\documentclass[graybox]{svmult}

\usepackage{mathptmx}       
\usepackage{amssymb}
\usepackage{helvet}         
\usepackage{courier}        
\usepackage{type1cm}        
\usepackage{makeidx}         
\usepackage{graphicx}        
\usepackage{multicol}        
\usepackage[bottom]{footmisc}

\makeindex             

\begin{document}

\title*{Cosmic rays and molecular clouds}
\author{Stefano Gabici}
\institute{Stefano Gabici \at AstroParticule et Cosmologie, Universit\'e Paris Diderot, CNRS/IN2P3, CEA/IRFU, Observatoire de Paris, Sorbonne Paris Cit\'e, 10, rue Alice Domon et L\'eonie Duquet, 75205 Paris Cedex 13, France, \email{Stefano.Gabici@apc.univ-paris7.fr}}
%
%
\maketitle

\abstract{This paper deals with the cosmic-ray penetration into molecular clouds and with the related gamma--ray emission. High energy cosmic rays interact with the dense gas and produce neutral pions which in turn decay into two gamma rays. This makes molecular clouds potential sources of gamma rays, especially if they are located in the vicinity of a powerful accelerator that injects cosmic rays in the interstellar medium. The amplitude and duration in time of the cosmic--ray overdensity around a given source depend on how quickly cosmic rays diffuse in the turbulent galactic magnetic field. For these reasons, gamma-ray observations of molecular clouds can be used both to locate the sources of cosmic rays and to constrain the properties of cosmic-ray diffusion in the Galaxy.}

\section{Introduction: the supernova remnant paradigm for the origin of galactic cosmic rays}
\label{sec:intro}

Cosmic Rays (CRs) \cite{berezinskii,gaisser,schlickeiser,longair} are charged and energetic particles that hit the Earth's atmosphere from above. The flux of CRs, once corrected for the effect of solar modulation, is constant in time and corresponds to a local energy density of $w_{CR} \approx 1$~eV/cm$^3$. Remarkably, this is comparable to the energy density of both magnetic field and thermal gas in the interstellar medium. CRs are mainly protons, with approximately 10\% of Helium, and 1\% of both heavier nuclei and electrons. Their differential energy spectrum is a steep and featureless power law $N_{CR} \propto E^{-s}$ with slope $s \approx 2.7$, and $\approx$~GeV particles are the main contributors to the total CR energy density. The slope of the spectrum slightly steepens  to $s \approx 3$ at an energy of $\approx 4 \times 10^{15}$~eV and this spectral feature is called the CR knee. The CR spectrum continues up to energies of the order of $\approx 10^{20}$~eV, but here we restrict ourselves to considering particles with energies below the knee, which can be confined by the interstellar magnetic field and thus are certainly of galactic origin. Finally, the arrival directions of CRs are extremely isotropic in the sky. The isotropy is of the order of $\approx 10^{-3}$ for particle energies  above $\approx$~1~TeV, where local (i.e. heliospheric) effects can be neglected, and depends very weakly on the particle energy for energies up to the knee. The high level of isotropy is due to the diffusion of CRs in the turbulent galactic magnetic field, which isotropizes the trajectories of particles and prevents a direct identification of CR sources based on the observed arrival direction of particles. This is the reason why indirect observational evidences, such as the detection of photons produced by CR interactions with the ambient medium, are needed in order to locate the sites of CR acceleration. {\it To date, the sources of galactic CRs are still not firmly identified.}

A connection between CRs and supernovae was first proposed by Baade and Zwicky in 1934 \cite{baadezwicky} and still remains the most popular explanation for the origin of galactic CRs. In its modern version (see \cite{hillas} for a review), the supernova paradigm for the origin of cosmic rays mainly relies on a consideration based on  the energy required to maintain the observed flux of CRs against their escape from the Galaxy. The overabundance of Li, Be, and B in CRs with respect to the abundances measured in the solar system, where they are virtually absent\footnote{Li, Be, and B are not synthesized in stars, and their standard abundance is very low, being mainly determined by primordial nucleosynthesis.}, can be explained as the result of spallation of heavier CR nuclei by interstellar gas. The amount of matter, or grammage, that CRs with an energy of $\gtrsim$GeV need to traverse to produce the observed amount of Li, Be, and B is equal to $\mu \approx 5$~g/cm$^2$. This corresponds to a confinement time in the galactic disk of $t_{d} = \mu/\rho c \approx 3 \times 10^6$~yr, where $\rho$ is the mean gas density in the disk ($\approx$~1 particle per cubic centimeter) and $c$ is the speed of light. Assuming that the CR intensity is constant in both time and space within the galactic disk, which has a radius of $R_{mw} \approx$~15~kpc and a thickness $h$ of a few hundred parsecs, one can estimate the CR luminosity of the Galaxy as $W_{CR} = [w_{CR} (\pi R_{mw}^2) h]/t_d \approx 10^{41}$~erg/s. This has to be compared with the total power from supernova explosions in the galaxy $P_{SN} = \nu_{SN} E_{SN} \approx 10^{42}$~erg/s, where $\nu_{SN} \approx 3$/century is the supernova rate in the Galaxy and $E_{SN} \approx 10^{51}$~erg is the typical supernova explosion energy. It is evident form these figures that {\it supernovae, or something related to them, may be the sources of CRs if $\approx$~10\% of their explosion energy is somehow converted into accelerated particles.}

A mechanism for the acceleration of particles that operates at supernova remnant (SNR) shocks was proposed in the late seventies, when it was realized that particles can be accelerated at shock waves via a first-order Fermi mechanism \cite{krimsky,bell78,blandfordostriker}.
A characteristic prediction of these models is a differential energy spectrum for the accelerated particles which is a power law with slope close to $\approx E^{-2}$.  
Power law spectra of relativistic particles have indeed been observed in SNRs, both through X-ray (see e.g. \cite{koyama,reynolds}) and gamma-ray (e.g. \cite{felixreview,jimreview}) observations and this is considered an evident manifestation of shock acceleration at expanding SNR shock waves.
The X-ray emission is unambiguously interpreted as the synchrotron radiation from relativistic electron, while the gamma ray emission can be interpreted either as inverse Compton scattering of electrons or decay of neutral pions generated in hadronic interactions between CRs and ambient gas. As discussed in the following, {\it the ambiguity between the hadronic or leptonic origin of the observed gamma-ray emission from SNRs is one of the main obstacles in proving (or disproving) the fact that SNRs are indeed the sources of CRs.}

Another issue that needs to be explained is the absence of features in the CR spectrum up to the energy of the CR knee, which is the mild steepening of the spectral slope observed at a particle energy of a few PeVs. The featureless of the spectrum up to that energy suggests that the sources that are responsible for the acceleration of the bulk of the CRs in the Galaxy ($\approx$GeV particles) are probably able to accelerate particles all the way up to the knee. In fact, assuming that several classes of sources contribute significantly to the CR spectrum at different particle energies and create such a featureless power law spectrum, though not impossible, would require an {\it ad hoc} fine tuning which seems quite unreasonable. In other words, {\it if SNRs are the sources of galactic CRs, most likely they have to act as particle PeVatrons} (see e.g. \cite{gabiciPeV}).

Finally, the question on the origin of galactic CRs cannot be considered answered until we understand the details of their propagation in the interstellar medium (for recent reviews see e.g. \cite{andyreview,fiorenzareview}). Measurements of the grammage that CRs must traverse while propagating from the sources to the Earth can be performed at different CR particle energies. Such measurements clearly point toward an energy dependent  grammage, and thus an energy dependent confinement time of CRs in the Galaxy, with higher energy particles escaping faster, according to $t_{esc}(E) \propto E^{-\delta}$, with $\delta \approx 0.3...0.6$. By assuming that CRs of all energies travel, on average, a distance $h$ before leaving the Galaxy, the escape time can be converted into a spatial diffusion coefficient $D \approx h^2/t_{esc} \approx D_0 (E/10~{\rm GeV})^{\delta}$, with $D_0 \approx 10^{28}...10^{29}$~cm$^2$/s \cite{andyreview,fiorenzareview}. If CR sources inject in the Galaxy $Q_{CR}(E)$ particles with energy $E$ per unit time, with a power law spectrum $Q_{CR}(E) \propto E^{-\alpha}$, then the equilibrium spectrum of CRs in the Galaxy is: $N_{CR} \propto Q_{CR}(E) \times t_{esc} \propto E^{-\alpha-\delta}$. The observed slope of the CR spectrum is $N_{CR} \propto E^{-2.7}$, which gives: $\alpha \approx 2.1...2.4$.
Thus, {\it the slope of the injection spectrum of CRs in the Galaxy has to be close to, but definitely steeper than 2}. 

At this point, another remark is needed. If the diffusion coefficient grows too fast with energy, i.e. if $\delta$ is closer to $\approx$~0.6 rather than to $\approx 0.3$, CRs with energies close to the knee would escape the Galaxy too quickly to be isotropized by the galactic magnetic field. In this scenario one would expect a high level of anisotropy at high energies, in contrast with what is observed. Thus, best--bet reference values for the spectral slope at injection of CRs and for the slope of the diffusion coefficient are probably $\alpha \approx 2.4$ and $\delta \approx 0.3$, which are consistent with both  the chemical and isotopic abundances of CRs and their isotropy.

The reason why the SNR hypothesis for the origin of galactic CRs is the most trusted and investigated scenario (but see e.g. \cite{bubbles} and \cite{dar} for a different and radically different perspective, respectively) is the fact that within this framework, most of the observational requirements can be explained within a reasonable accuracy.
As said above, the total energy budget is not an issue, provided that the efficiency of particle acceleration is of the order of $\approx$10\%. X-ray and gamma-ray observations clearly show that a mechanism capable of accelerating particles up to (at least) hundreds of TeVs operates at SNR shocks, and the characteristics of the observed radiation fit quite well with predictions of shock acceleration theory.
Moreover, recent developments in our understanding of the CR-induced amplification of the magnetic field at shocks suggest that SNRs might be able to accelerate particles up to the energy of the knee \cite{bell04} and inject them in the interstellar medium with a spectrum slightly steeper than $E^{-2}$ (see e.g. \cite{brianperp} or \cite{alfvendrift,driftcaprioli} for two ways to steepen the spectral slope above $\alpha = 2$), as required to explain the observed spectrum of CRs.
Also the CR chemical composition is reproduced with fair agreement with observations \cite{pzs}, while for what concerns the CR anisotropy the agreement between predictions and data is consistent within a factor of a few \cite{ptuskinanisotropy,pasqualeanisotropy}, if a weak dependence on energy of the diffusion coefficient is adopted, i.e. $\delta \approx 0.3$. However, in this latter case a comparison is less straightforward given that the level of anisotropy may be dominated by the exact location of the few nearest CR sources.

Despite all these very encouraging facts, it has to be kept in mind that we are still missing a conclusive and unambiguous proof of the fact that SNRs, as a class of objects, accelerate CRs and inject them in the interstellar medium at the rate required by observations. 
This review is an attempt to describe how gamma-ray observations, and in particular gamma-ray observations of molecular clouds, might finally lead to prove (or falsify) the SNR paradigm for the origin of CRs.

\section{Gamma-rays from supernova remnants}

CRs in the Galaxy undergo proton--proton hadronic interactions with the interstellar gas and produce neutral pions. The threshold energy for pion production is $E_p \approx 280$~MeV, where $E_p$ is the kinetic energy of the incoming CR. Each pion then decays into two gamma-ray photons which, for energies well above the threshold, have a typical energy equal to $\approx 0.1 \times E_p$ \cite{dermer,kelner,kamae}:
\begin{eqnarray}
p + p & \longrightarrow & p + p + \pi^0    \nonumber
\\
\pi^0 & \longrightarrow & \gamma + \gamma   \nonumber
\end{eqnarray}
Being the product of the decay of a particle of mass $m_{\pi^0}$, the resulting differential energy spectrum of gamma rays exhibits a pronounced peak at an energy $\approx m_{\pi^0}/2 \approx 70$~MeV. At energies larger than that, the gamma ray spectrum roughly mimics the spectrum of the CRs.   
The energy loss time of CRs due to proton-proton interactions is determined by the interaction cross section $\sigma_{pp} \approx 40$~mb and inelasticity $\kappa \approx 0.45$. Since these quantities depend weakly on particle energy, the energy loss time virtually depends on the gas density only and reads (e.g. \cite{atoyan}):
\begin{equation}
\label{eq:pp}
\tau_{pp} = \frac{1}{n_{gas} c \kappa \sigma_{pp}} \approx 6 \times 10^7 \left( \frac{n_{gas}}{1~{\rm cm}^{-3}} \right)^{-1} ~ {\rm yr}
\end{equation}
Charged pions can also be generated in proton--proton interactions, and the final products of their decay are electrons, positrons, neutrinos, and antineutrinos. Electrons and positrons can in turn produce gamma rays via inverse Compton scattering or Bremsstrahlung. Finally, also CR electrons, often referred to as {\it primary electrons}, to distinguish them from the ones ({\it secondaries}) generated in proton--proton interactions, can produce gamma rays through the same mechanisms.
Among all these mechanisms for the production of gamma rays, the decay of neutral pions is the dominant one in producing the prominent diffuse gamma-ray emission observed in the GeV energy range from the galactic disk \cite{EGRETdiffuse,FERMIdiffuse}. 
Such a strong diffuse emission constitutes an unavoidable background in searches for galactic sources of  GeV gamma-rays. On the other hand, due to the steepness of the CR spectrum, the diffuse TeV emission from the galactic disk is expected to be too weak to become an issue for present day Cherenkov telescopes, and for this reason in the following we focus mainly on the TeV energy domain.

A way to use gamma-ray observations in order to test the SNR hypothesis for the origin of CRs was proposed in 1994 \cite{dav,naito}. If SNR are the sources of CRs, $\approx 10$\% of the explosion energy of each supernova in the Galaxy has to be converted, on average, into CRs. The typical explosion energy is $\approx 10^{51}$~erg, and it does not vary much from supernova to supernova. Thus, on average one might expect to find $W_{CR}^{tot} \approx 10^{50}$~erg in form of CRs in a SNR. For simplicity, let's assume that at the SNR shock particles are accelerated with a differential spectrum $N_{SNR} \propto E^{-2}$ which extends from $\approx$~GeV to $\approx$~PeV energies. This gives a CR energy spectrum inside the SNR equal to $W_{CR} \approx N_{SNR}(E) E^2  \approx 7 \times 10^{48}$~erg. Then the gamma-ray flux from the SNR due to neutral pion decay is also a power law in energy with slope $\propto E_{\gamma}^{-2}$ and can be roughly estimated in this way:
\begin{equation}
\label{eq:flux}
F_{\gamma}(E_{\gamma})E_{\gamma}^2 \approx \frac{W_{CR} ~ c_{p\rightarrow\gamma}}{\tau_{pp} ~ (4 \pi d^2)} \approx 10^{-11} \left( \frac{W_{CR}^{tot}}{10^{50}{\rm erg}} \right) \left( \frac{n_{gas}}{1~{\rm cm}^{-3}} \right) \left( \frac{d}{1~{\rm kpc}} \right)^{-2} ~ {\rm erg/cm^2/s}
\end{equation}
where $c_{p\rightarrow\gamma}\approx 0.1$ is the average fraction of the proton energy transferred to the gamma-ray photon and $d$ is the SNR distance. In computing the gamma-ray flux it has been assumed that the ambient gas with density $n_{gas}$ is compressed at the strong SNR shock by a factor of four. Moreover, any leptonic contribution to the gamma-ray emission (i.e. inverse Compton scattering and Bremsstrhalung from both primary and secondary electrons) has been neglected. 

The hadronic gamma-ray flux predicted from Eq.~\ref{eq:flux} is well within the detection capability of the Cherenkov telescopes that currently operate in the TeV energy domain:  H.E.S.S., MAGIC, and VERITAS. In other words, if SNR are the sources of CRs, some of them must have been detected by Cherenkov telescopes.
This claim remains substantially correct, thought less striking \cite{pierre}, even if one assumes a steeper spectrum $\approx E^{-2.1...2.4}$ for the accelerated particles, a choice more consistent with CR data (see Sec.~\ref{sec:intro}). 
Indeed, roughly half a dozen of isolated (i.e. not associated with a molecular cloud) SNRs have been detected in TeV gamma rays (for reviews see \cite{felixreview,jimreview}), and this fits well with the above mentioned predictions. However, electrons are also accelerated at shocks, as demonstrated by the X-ray synchrotron emission observed from a number of SNRs. Thus, a competing leptonic mechanisms, namely inverse Compton scattering off photons of the cosmic microwave background radiation, could also account for the observed TeV emission. This means that the detection of SNRs at TeV energies cannot be considered as a proof of the fact that SNRs are indeed the sources of CRs, but only as an additional consistency check of that scenario.

The question about the hadronic or leptonic nature of the TeV emission from SNRs constitutes one of the most discussed issues in gamma-ray astronomy. 
A key parameter which regulates the predominancy of one contribution over the other is the magnetic field in the shock region.
If the field is significantly stronger than $\approx 10 ~ \mu$G, then the observed
synchrotron X-rays can be explained by a relatively meagre number of electrons,
which would produce unappreciable TeV inverse Compton emission. Conversely, if
the value of the magnetic field is $\gg 10 ~ \mu$G the much larger number of electrons needed to explain the X-ray emission will also suffice to explain the whole observed TeV emission. Thus, the value of the magnetic field at the shock is a crucial parameter of the problem, and its determination would allow us to unveil the nature of the gamma ray emission.

Observational evidence has been found for the presence of very strong magnetic fields of the order of $\approx 100~\mu$G...1~mG in several  SNRs. Such evidences come from X-ray observations of  narrow synchrotron filaments located at the position of the shock \cite{aya,jacco,heinzfilaments}, or of very fast temporal variations of the synchrotron emission from small knots within the SNR \cite{yas1,yas2}. In both cases, a strong magnetic field is needed to reduce the synchrotron cooling time of electrons down to values that would explain both the fast variability and the small thickness of filaments, since in such a strong magnetic field electrons would radiate all their energy before reaching large distances downstream of the shock. The importance of these measurements is the fact that they point towards a very efficient acceleration of CR protons and nuclei, which seem to be the only plausible source for the field amplification. On the other hand, it has to be noted that the regions in which field amplification has been measured constitute a very small fraction of the entire SNR shell, and thus little can be said about the average value of the field on larger scales, and also on the global CR acceleration efficiency from the whole shock\footnote{At least one exception to this exists, which is the SNR Cas~A, for which a high value ($\approx 300 ~ \mu$G) of the average magnetic field in the whole shell region has been determined \cite{cowsiksarkar,atoyanCasA}.}. In other words, the average field could be much smaller than the one measured in knots and filaments, and this would be in agreement with a leptonic interpretation of the gamma-ray emission from SNRs.

Another way to discriminate between the hadronic or leptonic origin of the observed gamma-ray emission from SNRs is by measuring the spectral slope of the gamma-ray spectrum from the GeV to the TeV energy range.  SNR shocks are expected to accelerate particles, both hadrons and leptons, with a differential spectrum which is close to a power law $N(E) \propto E^{-\gamma}$. By using a very simple toy--model it can be shown that the spectrum of the CRs that escape the SNRs and are injected in the interstellar medium is expected to be, very roughly, $\approx E^{-2}$ if $\gamma < 2$ and $\approx E^{-\gamma}$ if $\gamma > 2$ \cite{ptuskinzirakashvili,ohiraescape,damianoescape,gabiciescape}. Thus, to satisfy the requirements from CR data, we adopt here values of $\gamma$ in the range 2.1...2.4. This would result in a relatively steep (i.e. steeper than $E^{-2}$) hadronic gamma-ray emission due to proton-proton interactions with a spectrum $E_{\gamma}^2 F_{\gamma}^{pp}(E_{\gamma}) \propto E_{\gamma}^2 N(E/c_{p\rightarrow\gamma}) \approx E_{\gamma}^{-0.1...0.4}$. To compute the spectrum resulting from the inverse Compton scattering of accelerated electrons one has to keep in mind that in the leptonic scenario for the gamma--ray emission a weak magnetic field is required in order not to overshoot the observed X--ray fluxes. Thus, synchrotron energy losses can be neglected and the spectrum of electrons does not change and remains $E^{-\gamma}$. In this case, the differential energy spectrum of the inverse Compton scattering emission is a power law with index $(\gamma+1)/2 \approx 1.55...1.7$, resulting in a spectral energy distribution equal to $E_{\gamma}^2 F_{\gamma}^{IC} \approx E_{\gamma}^{0.3...0.45}$, which is significantly harder than $E^{-2}$. This spectral difference -- a hard spectrum for the leptonic emission and a soft one for the hadronic -- can be used to distinguish between the two scenarios. To do this, GeV observations by the Fermi satellite are of crucial importance since they can be combined with TeV observations to obtain a broad band gamma--ray spectrum.

Indeed, Fermi detected several young SNRs. Observations of the SNR RX~J1713.7-3946, the most prominent TeV bright SNR, revealed a very hard spectral energy distribution with index close to 1.5, suggesting that for this particular SNR the gamma-ray emission is most likely leptonic \cite{fermiRXJ}. Conversely, the gamma-ray spectrum observed from the historical SNR Tycho is a power law with slope $\approx 2.3$, pointing towards an hadronic origin of the emission \cite{fermiTycho,giovannidamiano}. The case of Vela Jr., another TeV bright young SNR, is less clear, since neither the hadronic nor the leptonic model can be ruled out based on gamma--ray data only \cite{velaJrFermi}. Similar conclusions can be reached for the SNR Cas A, for which an hadronic origin of the gamma--ray emission seems favored, but a leptonic one (based on relativistic Bremsstrahlung) cannot be completely ruled out \cite{FermiCasA}. To complete the picture, some remarks are in order: first of all, the fact that the emission from RX~J1713.7-3946 seems to be leptonic does not mean that that SNR is not an accelerator of CR protons. In fact, if the ambient density close to the SNR is low, at the level of $\approx 0.1$~cm$^{-3}$, then the expected hadronic gamma--ray emission would fall below the observed one even for quite high acceleration efficiency of $\approx 30\%$ \cite{fermiRXJ}. Thus, the evidence for leptonic emission cannot be considered as an argument against CR acceleration at SNRs. The second remark is that a hadronic interpretation for the emission from RX~J1713.7-3946 has been pushed forward in \cite{fukuiRXJ,fukuiRXJ2}, while a leptonic one for Tycho has been presented in \cite{dermerTycho}. This indicates that an unanimous agreement on the interpretation of the gamma--ray emission from young SNRs is not reached yet, even for the best studied sources.

X--ray observations can also help in discriminating between hadronic and leptonic models. In fact, the first convincing evidence in favor of the leptonic origin of the gamma rays from the SNR RX~J1713.7-3946 came from the non-detection of X--ray lines in its spectrum. The idea is that if one wants to explain the gamma--ray emission as the result of hadronic interactions, a high density for the ambient gas has to be assumed. This would also enhance the thermal emission from the shocked gas, and X--ray lines should appear on top of the synchrotron X--ray spectrum. Their absence in the observed spectrum cannot be accounted for in hadronic models \cite{donRXJ}, unless two distinct zones (one for the production of gamma--rays and one for the production of thermal radiation) are invoked \cite{fukuiRXJ}.

Finally, Fermi also detected several old SNRs, with an age of $\approx 10^4$~yr (e.g. \cite{FermiW51,FermiW44}). In most cases the gamma--ray emission is likely to have a hadronic origin. This is because a leptonic interpretation would require an unreasonable total energy in form of accelerated electrons (e.g. \cite{FermiW51}). For old SNRs, the velocity of the shock is quite small, of the order of 100 km/s, and these objects are very often interacting with dense molecular clouds. Both these things may inhibit the acceleration (or reacceleration \cite{crushedyas}) of particles. On one side, the acceleration rate is small for slow shocks and in addition to that, the neutral gas in the molecular cloud can damp the magnetic field turbulence and reduce the scattering rate of particles, which is a key ingredient for their acceleration \cite{malkovbreak}. Thus, old SNRs are not expected to accelerate CRs all the way up to the knee and, in agreement with this expectation, their TeV gamma--ray spectrum, when measured, is generally quite steep.

To conclude, there is now growing consensus on the fact that SNRs are capable of accelerating CR protons up to GeV--TeV energies. Also, there is quite convincing indication for proton acceleration above TeV energies from at least one young SNR (Tycho), while in another case (RX~J1713.7-3946) the gamma--ray emission is most likely leptonic, and we cannot say much on CR acceleration efficiency from gamma--ray data only. Moreover, the high values of the magnetic field inferred from X-ray observations of several young SNRs seem to require that efficient CR acceleration operates at least at some locations in those shocks. Unfortunately, we are still missing convincing and direct observational evidence for the fact that SNRs accelerate particles up to the knee. A conclusive proof (or confutation) of the SNR paradigm for the origin of CRs will hopefully come soon, with the advent of new gamma--ray facilities like CTA, HAWK or LHAASO \cite{CTA,HAWK,LHAASO}, which will dramatically increase the amount and quality of data. 

In the next sections, it will be shown how present and future gamma--ray observations of molecular clouds can contribute in solving the problem of CR origin. In particular, if a massive cloud is located at or close to the SNR shock, the gamma--ray emission due to neutral pion decay is strongly enhanced due to the presence of a thick target. This would enhance the probability to detect the hadronic emission and thus allow the identification of SNRs as sources of CRs. Moreover, in at least two cases, the SNRs W28 \cite{W28HESS,W28FERMI} and W44 \cite{yasW44}, some gamma--ray emission is detected from outside of the SNR shell, and coincident with the position of dense gas clouds. In another case, the SNR IC~443, the centroids of the GeV and TeV emissions are not coincident, but significantly displaced \cite{FermiIC443}. To interpret these observations, CR escape from the SNR shells and energy dependent propagation of CRs have been often invoked. This suggests that gamma--ray observations of molecular clouds could also be used to study and constrain the propagation of CRs close to their sources, about which very little is known.

\section{Molecular clouds as cosmic ray barometers}

Consider a molecular cloud (MC) with mass $M_5 = (M/10^5 M_{\odot})$ at a distance $d_{\rm kpc} = (d/{\rm kpc})$ from the observer. Let us further assume that the CR intensity in the region of the Galaxy where the MC is located is the same as the one measured at the Earth. Under these circumstances, an expression similar to Eq.~\ref{eq:flux} can be written to describe the expected integral gamma-ray flux due to proton--proton interactions in the MC \cite{blackfazio,felixclouds}:
\begin{equation}
\label{eq:cloud}
F_{\gamma}(> E_{\gamma}) \approx 2 \times 10^{-13} ~ \delta ~ \left( \frac{M_5}{d_{\rm kpc}^2} \right)  \left( \frac{E_{\gamma}}{\rm TeV} \right)^{-1.7}  {\rm cm^{-2} s^{-1}}
\end{equation}
where a multiplicative factor of $\approx 1.5$ has been applied to account for the contribution to the emission from nuclei heavier than hydrogen both in CRs and ambient gas \cite{mori}. The factor $\delta$ accounts for possible deviation of the CR intensity with respect to the one measured at the Earth, and under the assumption (quite reasonable, within a factor of a few) of homogeneity of the CR intensity in the Galaxy it is equal to 1. By assuming that the MC has an average density of $\approx 100$~cm$^{-3}$ one can estimate its apparent size as: $\vartheta_{cl} \approx 1^{\circ} M_5^{1/3}/d_{\rm kpc}$. MCs characterized by $\delta = 1$ (i.e. no CR overdensity with respect to the CR background) have been often referred to as {\it passive} clouds \cite{passive}, to indicate the absence of particle acceleration inside or in the vicinity of the cloud.

From the discussion above it follows that: {\it i)} MCs are expected to be quite extended TeV gamma--ray sources, since Cherenkov telescopes have an angular resolution of $\vartheta_{res} \approx 0.1^{\circ}$, {\it ii)} Cherenkov telescopes of current generation, with an integral sensitivity for point sources of the order of $\Phi(> 1~{\rm TeV}) \approx 10^{-12}$~cm$^{-2}$~s$^{-1}$, which worsen as $\approx \vartheta_{cl}/\vartheta_{res}$ for extended sources, cannot detect passive MCs unless they are much more massive than $10^5 M_{\odot}$ and/or very close to the Earth, {\it iii)} the future TeV gamma--ray facility CTA, the Cherenkov Telescope Array, will improve the sensitivity by a factor of 5-10 and will be able to detect passive clouds with masses $\gtrsim 10^5 M_{\odot}$ only if they are located within a distance of $\lesssim 1$~kpc or so \cite{meCTA}.

If the assumption $\delta = 1$ is relaxed, and gamma rays are detected from a MC, Eq.~\ref{eq:cloud} can be used to determine the actual value of $\delta$ at the location of the cloud, provided that the mass and distance of the MC are known. If not only the intensity of CRs, but also their spectral distribution differs from the local one, $\delta$ becomes an energy dependent quantity. Overdensities of CRs of the order of $\delta > 10$, or masses well above $10^5 M_{\odot}$ are needed in order to detect TeV photons from MCs with telescopes of present generation. This is illustrated with a few examples below.

The molecular cloud complex located in the inner Galaxy and named {\it galactic centre ridge} has been detected by H.E.S.S. \cite{HESSridge} as a diffuse TeV emission extending for $\approx 1^{\circ}$ or so around the galactic centre. The spatial distribution of the gamma--ray emission correlates well with the gas density in the region, which is derived from the observations of the CS emission line \cite{CS}. This suggests that the emission is of hadronic origin, the dense gas being the target for CR interactions. A detection has been possible because the large distance of $\approx 8.5$~kpc is compensated by the very large gas mass of the complex, equal to a few times $10^7 M_{\odot}$. The differential spectrum of the gamma--ray emission can be fitted by a power lax with index $\approx 2.3$, which indicates that the spectrum of the CRs responsible for that gamma--ray emission is much harder than the CR galactic background, which is $\approx E^{-2.7}$. Moreover, from the measured gamma--ray flux at 1 TeV it can be inferred that the intensity of $\approx 10$~TeV CRs in the galactic centre region is enhanced by a factor of $\approx 3...10$ with respect to the background. The hard spectrum, coupled with the enhanced intensity of CRs implies that some recent event of particle acceleration took place in that region.

Another similar example is represented by the massive MCs located in the vicinity of the SNR W28 \cite{W28HESS}. These clouds have a total mass of $\approx 10^5 M_{\odot}$ and the system (SNR and MCs) is located at a distance of $\approx 2$~kpc. The MCs have been detected in TeV gamma rays by H.E.S.S. and a CR overdensity of the order of $\delta \approx 10...30$ has been inferred. As it will be discussed in Sec.~\ref{sec:W28}, the most natural interpretation of that emission is that the CRs in excess of the background have been accelerated by the SNR W28, have escaped the site of acceleration and are still diffusively confined in the region. The enhanced intensity of CRs coupled with the presence of the dense clouds can explain the observed gamma--ray emission. 

In the GeV energy domain these kind of studies are more difficult because of the worse angular resolution (a fraction of a degree) of current instruments and most of all because of the very intense diffuse emission from the galactic disk that constitutes an important background in the search of gamma--ray sources. However, studies of the diffuse emission and of prominent clouds or cloud complexes remain feasible, and could provide information on the large scale distribution of CRs (see e.g. \cite{sabrinaclouds}). The Fermi collaboration performed a study of the diffuse emission from some specific regions in the galactic plane characterized by the presence of dense gas, like the Cassiopeia and Cepheus region \cite{fermicepheus} and the third galactic quadrant, where the Local and Perseus arms can be observed \cite{fermithird}. From the knowledge of the spatial distribution of the gas along the line of sight, the spatial distribution of $\approx$~GeV CRs has been inferred. No significant spectral variation up to the outer Galaxy has been found, meaning that the spectrum of $\approx$~GeV CRs is, on large galactic scales, quite constant. The CR intensity has been found to have a weak dependence on galactocentric distance, even weaker than the one predicted by standard propagation models such as GALPROP \cite{GALPROP}. This means that the CR intensity is not expected to vary a lot (i.e. more than a factor of a few) over galactic scales. 

It is clear from the examples described above how gamma--ray observations of MCs can be used to measure the CR intensity in specific regions of the Galaxy. For this reason, MCs have been sometimes referred to as {\it CR barometers} and are now considered important tools to locate the position of galactic accelerators of CRs (e.g. \cite{snobs,issa,atoyan,gabiciPeV,diegoolaf,sabrinaclouds}). In fact, the importance of the gamma--ray emission from MCs was first realized in connection to the estimate of the masses of MCs \cite{blackfazio}. These masses are estimated from the intensity of emission lines of some molecules, most notably the carbon monoxide, CO, which is the second most common molecule in the interstellar medium after molecular hydrogen, and it is much easier to detect. The line intensity gives the mass of the CO in the cloud, which can be converted to the total mass, largely dominated by molecular hydrogen, through a conversion factor, $X_{CO}$, which unfortunately is not very well constrained \cite{combes}. The idea to use gamma--rays to estimate the mass of MCs is based on the assumption of spatial homogeneity of CRs in the Galaxy, which would allow to use the gamma--ray flux observed from MCs to calibrate the $X_{CO}$ conversion factor. This can be easily understood from Eq.~\ref{eq:cloud} with $\delta = 1$. Very recently, the Fermi collaboration performed a study of the massive and nearby Orion MCs \cite{fermiorion} and obtained interesting constraints on the value of $X_{CO}$, pointing towards a non--linear relation between the densities of carbon monoxide and molecular hydrogen.

Thus, it is clear from what said above that gamma--ray bright MCs can be used as CR barometers with one main caveat: the uncertainty on the value of $X_{CO}$, which is used to determine the mass of the cloud, translates in an uncertainty on the value inferred for the CR overdensity $\delta$ from the gamma--ray flux of the cloud. Despite that, MCs can still be used to measure variations in the CR intensity in the Galaxy. First of all, spectral measurements are not affected by the uncertainty in the determination of the mass. As an example, the hard gamma--ray spectrum observed from the galactic centre ridge implies that the CRs that produce that gamma--ray emission have a different spectrum, and thus intensity, with respect to the galactic CR background. Second, large overdensities of CRs of the order of a few tens, as the one inferred for the clouds in the vicinity of the SNR W28, can hardly be ascribed solely to an erroneous determination of the mass of the cloud.

One last issue that needs to be discussed concerns the penetration of CRs into MCs. So far, it has been implicitly assumed that CRs can freely penetrate MCs and that the CR intensity inside a cloud is equal to the one immediately outside. In fact, this assumption is valid only when the time it takes CRs to diffuse across the cloud is shorter than the CR energy loss time (see e.g. \cite{gabici2007}).

\begin{figure}[t!]
 \centering
 \includegraphics[width=0.49\textwidth]{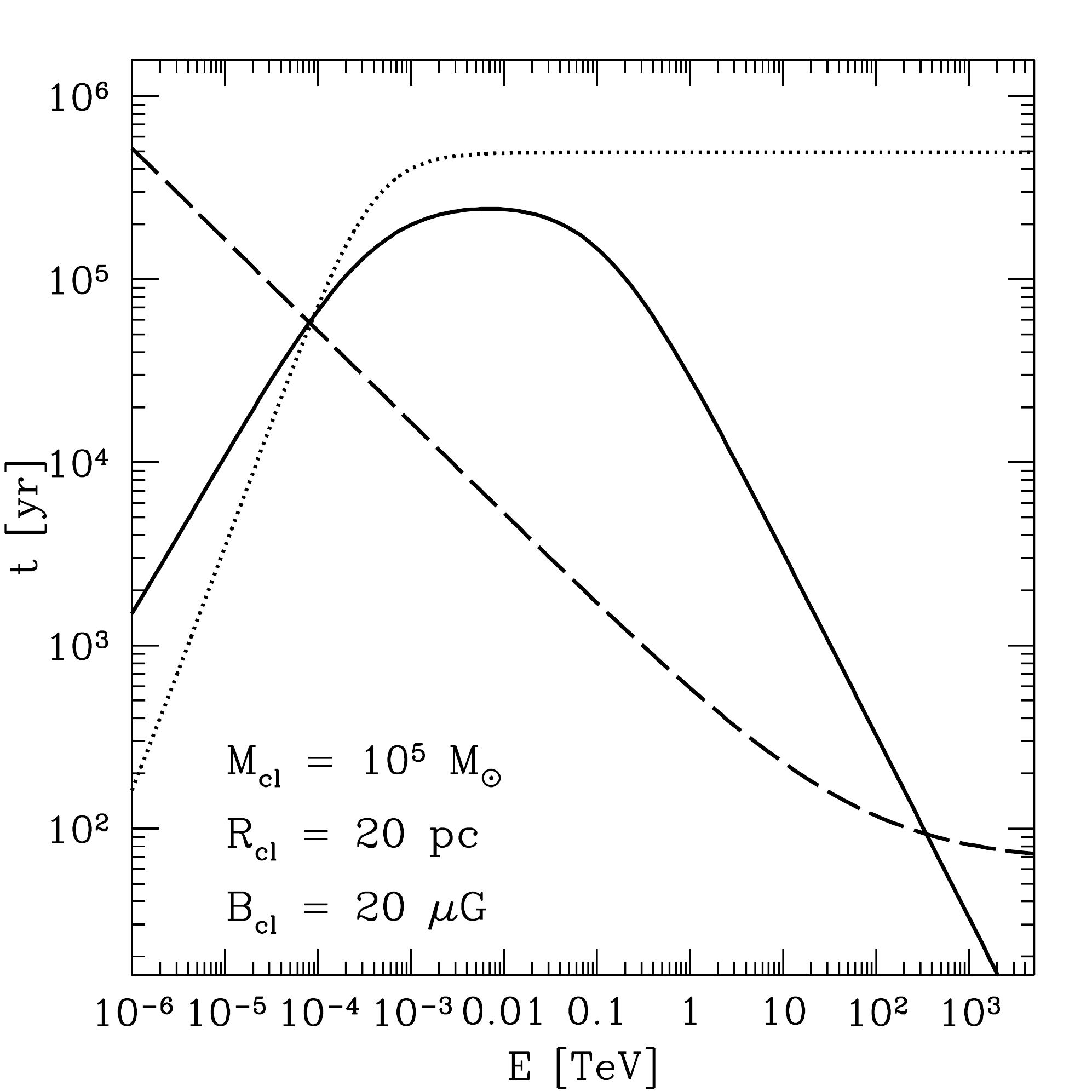}      
  \includegraphics[width=0.49\textwidth]{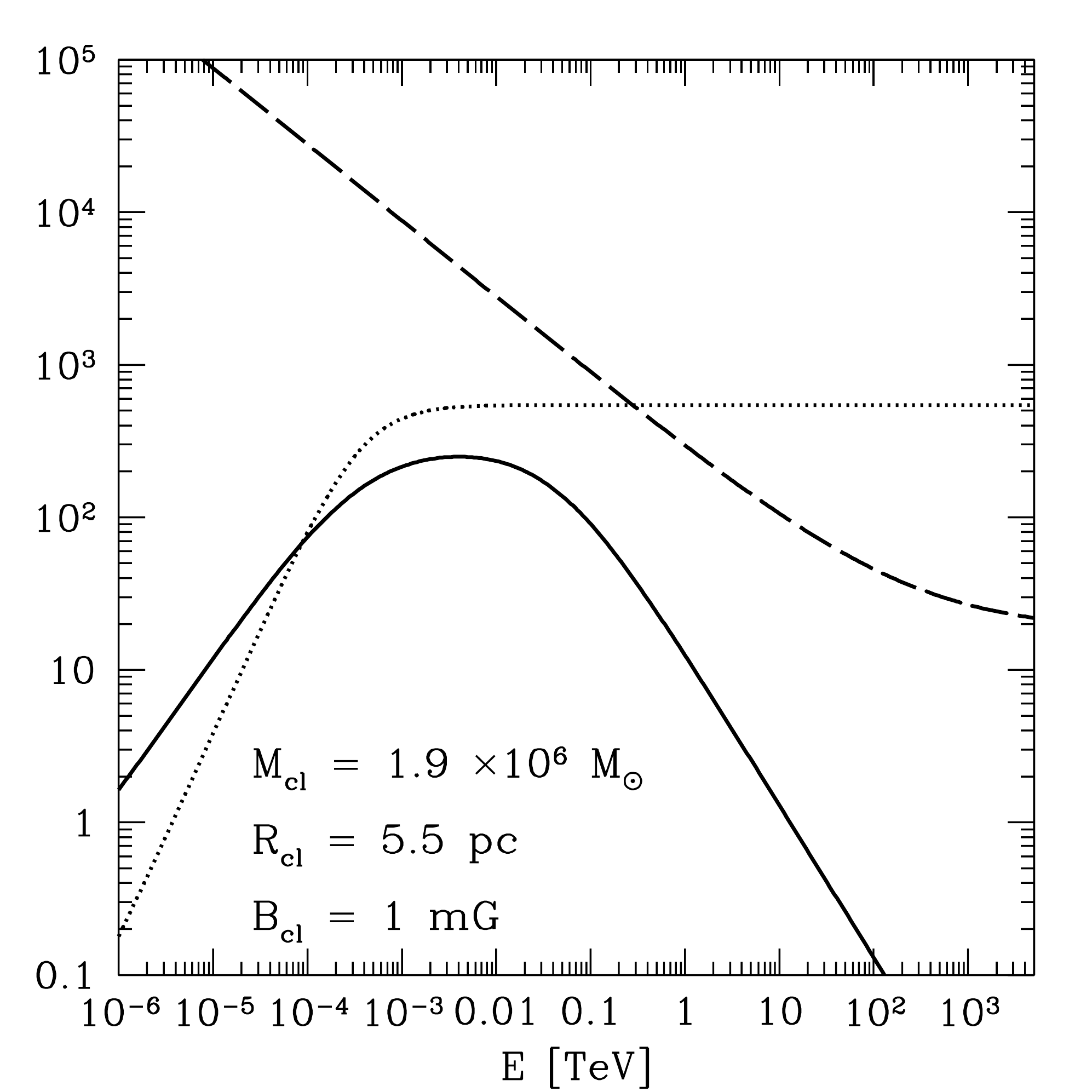}      
  \caption{{\bf LEFT PANEL:} relevant time scales for CR propagation inside a molecular cloud with mass $M_{cl} = 10^5 M_{\odot}$, radius $R_{cl} = 20$~pc, and magnetic field $20~\mu$G. Assuming a flat density profile the density is $\sim 120$~cm$^{-3}$. The dashed line represents the CR propagation time scale over a distance $R_{cl}$. The dotted line represents the energy loss time for CR protons (ionization losses are relevant below 1 GeV, inelastic proton-proton interaction at higher energies) while the solid line refers to the energy loss time for CR electrons, including ionization losses, Bremsstrahlung losses and synchrotron losses, which dominates at low, intermediate and high energies respectively. {\bf RIGHT PANEL:} same as the left panel, but for the specific case of the SgrB2 cloud with mass $1.9 \times 10^6 M_{\odot}$, radius 5.5 pc and magnetic field 1 mG. This implies an average density of $1.1 \times 10^5$ cm$^{-3}$ (see text for an explanation of the choice of parameters adopted). Figure from \cite{gabici2009}}
  \label{fig:timeGMC}
\end{figure}

To check whether this is true or not, in the left panel of Fig.~\ref{fig:timeGMC} \cite{gabici2009} the characteristic time scales for diffusion and energy losses have been plotted as a function of the particle energy for a giant molecular cloud with total mass $M_{cl} = 10^5 M_{\odot}$ and radius $R_{cl} = 20$ pc. Assuming a flat density profile the density is $n_{gas} \sim 120$ cm$^{-3}$. The magnetic field is assumed to be $B_{cl} = 20~ \mu$G. The dotted line refers to proton energy losses, which are dominated by ionization losses at energies below $\sim 1$ GeV and by inelastic proton--proton interactions at higher energies. The solid line represents the electron energy loss time. The three different power law behaviors reflect the dominance of ionization, Bremsstrahlung and synchrotron losses at low, intermediate and high energies, respectively. Finally, the dashed line represents the propagation time over a distance equal to the cloud radius. The propagation time has been estimated as $t_d \approx R_{cl}^2/(6~D)$ with a diffusion coefficient equal to $D = 10^{28} (E/10~{\rm GeV})^{0.5}~{\rm cm^2/s}$, a value consistent with CR data. The deviation from a power law behavior at high energies indicates the transition from a diffusive to a straight line propagation.

For proton energies above the threshold for pion production ($E_{th} \sim 280$ MeV), the propagation time is always shorter than the energy loss time. This means that CR protons which produce both gamma rays and secondary electrons can freely penetrate the cloud and their flux is not attenuated due to energy losses.
The propagation time for CR electrons is also shorter than the energy loss time for particle energies between $E \sim 100$ MeV and a few hundreds of TeV. This implies that, within this energy range, the secondary electrons produced inside the cloud quickly escape, and have little effect on the non-thermal emission from the cloud. On the other hand, extremely energetic electrons with energies above a few hundreds TeVs radiate all their energy in form of synchrotron photons before leaving the cloud. In a typical magnetic field of a few tens of microGauss, these electrons emit synchrotron photons with energy:
$
E_{syn} \approx 1 (B_{cl}/10~\mu {\rm G}) (E/100 ~ {\rm TeV})^2 ~ {\rm keV}  .
$
Thus, the most relevant contribution from secondary electrons to the cloud non thermal emission falls in the hard X-ray band. 


The properties of giant molecular clouds located in the galactic centre region can differ significantly from the average figures reported above.
As an example, we plot in the right panel of Fig.~\ref{fig:timeGMC} the typical time scales for the SgrB2 cloud.
This is a very massive cloud located at 100 pc (projected distance) from the galactic centre. 
The cloud virial mass is $M_{SgrB2} = 1.9 \times 10^6 M_{\odot}$ \cite{protheroe} and a magnetic field at the milliGauss level has been measured in the outer envelope of the cloud complex \cite{crutcher}.
The mass distribution can be fitted with a radial gaussian density profile with $\sigma = 2.75$ pc \cite{protheroe}.
To compute the curves plotted in Fig.~\ref{fig:timeGMC} (right panel), we assumed a cloud radius of $R_{SgrB2} = 2 \sigma = 5.5$~pc, which encloses $\approx 95\%$ of the total cloud mass. This gives an average density of $n_{gas} = 1.1 \times 10^5$cm$^{-3}$ (roughly a factor of 2 below the central density).
It is evident from the right panel of Fig.~\ref{fig:timeGMC} that the SgrB2 cloud is remarkably different from a typical giant molecular cloud.
In particular, the very high values of the magnetic field and of the gas density make the energy loss time of CR protons significantly shorter than the propagation time for energies below a few hundred GeVs.
Moreover, for CR electrons the energy loss time is always shorter than the propagation time.
This means that CR protons with energies up to few hundred GeVs cannot penetrate the molecular cloud, as they do in the cases considered in the left panel of Fig.~\ref{fig:timeGMC}.
Primary CR electrons cannot penetrate the cloud, while secondary CR electrons produced inside the cloud in hadronic interactions cannot leave the cloud and radiate all their energy close to their production site.
These characteristics make SgrB2 a very peculiar objects whose modeling needs a specific treatment.

A detailed study of the effects of CRs exclusion from giant MCs on their gamma ray emission have been discussed in detail in \cite{gabici2007} and the main results can be summarized as follows: CRs are expected to penetrate freely giant MCs unless the diffusion coefficient inside the cloud is substantially reduced (i.e. at least a factor of 100...1000) with respect to the average galactic one. An exception to this is the SgrB2 MC, characterized by an unusually large mass and density which would prevent CRs penetration due to strongly enhanced energy losses.

Now that the issue of the CR penetration inside MCs has been addressed, it is possible to discuss the situation in which a MC is located in the vicinity of a CR accelerator, and attempt to predict the gamma--ray spectrum resulting from the interactions between the CRs that escaped the accelerator and the dense gas in the cloud. As stressed above, the detection of such emission can be used to locate the sources of CRs, but also to study their propagation properties on spatial scales of the order of the distance between the CR source and the MC.

\section{Molecular clouds/supernova remnant associations: \\ expected  cosmic-ray and gamma-ray spectra}
\label{sec:SNRMC}

After escaping the acceleration site (e.g. a SNR shock), CRs interact with the ambient gas and produce neutral pions that in turn decay into gamma rays.
The production of such radiation is enhanced if a large amount of dense gas (e.g. a massive MC) is present in the vicinity of the source of CRs. The CR background in the Galaxy has a steep differential spectrum with slope $s \approx 2.7$ and an energy density equal to $E^2 N_{CR}(E) \approx 6 \times 10^{-3} (E/{\rm TeV})^{-0.7} {\rm eV~cm^{-3}}$. Due to the steepness of the spectrum, an excess in the CR intensity with respect to the CR background  would appear more easily at higher ($\approx$~TeV) than at lower ($\approx$~GeV) energies. For this reason we focus here on the energy domain probed by Cherenkov instruments, which corresponds to photon energies $\gtrsim 100$~GeV.

To be more specific, consider now a SNR which releases in a single impulsive event $\approx 10^{50}~W_{50}$~erg of CRs protons in the interstellar medium, with a differential spectrum $N_{SNR}(E) \propto E^{-\alpha}$. 
If we assume that SNRs indeed are the sources of galactic CRs, then we know from the constraints coming from CR data (see Sec.~\ref{sec:intro}) that, on average, $W_{50} \approx 1$ and $\alpha$ has to be in the range $\approx2.1...2.4$.
For example, for a CR spectrum with slope 2.4 this corresponds to a total differential spectrum equal to: $E^2 N_{SNR}(E) \approx 2 \times 10^{60} W_{50} (E/{\rm TeV})^{-0.4}$~eV. If CR diffusion proceeds isotropically, at a given time $t$ after escaping the SNR particles with a given energy $E$ occupy (roughly uniformly) a spherical region of radius $R \approx \sqrt{6 ~ D(E) ~ t}$ surrounding the SNR, where $D$ is the diffusion coefficient.
It follows that the energy density of runaway CRs with an energy of 1~TeV exceeds the one in the background if CRs have propagated up to a distance $R_* \lesssim 140 ~ W_{50}^{1/3}$pc from their source. If one takes $D_{gal} = 10^{29}$~cm$^2$/s as a reference value for the galactic diffusion coefficient of CRs with energy of 1 TeV, then the time on which an excess of CRs is present around the SNR is $t_* \approx R_*^2/(6~D_{gal}) \approx 10^4 ~ {\rm yr}$.
Therefore, the search for gamma ray emission from molecular clouds located close to SNRs has to be focused on regions of size $100-200$~pc around SNRs not much older than $\approx 10^4$~yrs.
These figures, together with the spectrum and the intensity of the gamma--ray emission, change if the diffusion coefficient in the vicinity of the SNR differs significantly from its average value $D_{gal}$, and this suggests that these kind of studies might be used to constrain the diffusion properties of CRs close to their sources. This is very important, because from CR data only a value of the diffusion coefficient averaged over a large volume in the Galaxy can be inferred, and little is known on its spatial variations on small scales. 

One reason to expect a smaller diffusion coefficient close to CR sources is connected to the presence of the CRs themselves. This is because CRs that escape the source can generate magnetic turbulence via streaming instability and such turbulence can in turn confine CRs, reducing the diffusion coefficient \cite{kulsrud,wentzel,cesarsky}. Under these circumstances, the CR diffusion becomes a non--linear process \cite{plesser,malkovfelix}. It follows that observational constraints on the diffusion coefficient can shed light on the plasma instabilities through which CRs generate magnetic turbulence.

Following \cite{atoyan}, we consider two different scenarios for the escape and propagation of CRs away from the source: in the first one CRs are released in the interstellar medium in a single impulsive event occurring at a time $t = 0$, while in the second one CRs are continuously released over a time interval $\Delta t$. In a rough approximation, these two scenarios describe two possible (and opposite) ways in which SNRs may release CRs in the interstellar medium: in an almost impulsive event at some stage of the dynamical evolution of the SNR, or continuously over an extended time interval (e.g. the whole Sedov phase). These two extreme and definitely idealized scenarios should account for our little knowledge on the actual way in which CRs escape SNRs \cite{gabiciescape}.  

In both scenarios, the propagation of CR protons in the interstellar medium can be described by the diffusion equation:
\begin{equation}
\label{eq:prop}
\frac{\partial f}{\partial t} = D ~ \nabla^2 f  + Q
\end{equation}
where $f(E,R,t)$ is the differential energy distribution of CRs at a given time $t$ and at a distance $R$ from the source,  $D(E) \propto E^{-\delta}$ is the CR diffusion coefficient, assumed to be isotropic and spatially homogeneous, and $Q(E,R,t)$ is the CR injection term. Energy losses have been neglected because, as can be estimated from Eq.~\ref{eq:pp}, for the typical densities of the interstellar medium the energy loss time far exceeds the CR confinement time in the Galaxy. Moreover, we do not consider here CR electrons because they are expected to suffer severe synchrotron losses at the SNR shock. Such strong energy losses might prevent them to escape the shock until the very late phases of the SNR evolution.

For an impulsive injection of CRs in the interstellar medium, the injection term reads: $Q = N_{SNR}(E) \delta(R) \delta(t)$, where we have assumed a point like source. Here, $N_{SNR}(E) \propto E^{-\alpha}$ represents the number of particles of energy $E$ which escape the SNR and is normalized to $\int {\rm d}E ~ N_{SNR}(E) ~ E = W_{CR}$, where $W_{CR} = 10^{50} ~ W_{50} ~ {\rm erg}$ is the total energy in form of escaping CRs. Under these assumptions the solution of Eq.~\ref{eq:prop} is \cite{atoyan}:
\begin{equation}
\label{eq:impulsive}
f(E,R,t) = \frac{N_{SNR}(E)}{\pi^{3/2} R_d^3} \exp \left[ -\left(\frac{R}{R_d}\right)^2 \right]
\end{equation}
where $R_d = \sqrt{4~D~t}$ is a characteristic diffusion distance, i.e. the distance that particles with energy $E$ can cover in a time $t$. Thus, if we consider a specific energy $E$, for distances smaller than $R_d$ Eq.~\ref{eq:impulsive} can be approximate, neglecting factors of order unity, as: $f \approx W_{CR}/R_d^3$ which means that CRs are distributed roughly homogeneously within a spherical region of radius $R_d$. If a MC is located at a given distance $d_{SNR/MC}$ from the SNR, by using the definition of the diffusion distance $R_d$ it is possible to estimate the minimum energy $E_*$ of CRs that can diffuse up to the cloud in a time $t$: $d_{SNR/MC} = \sqrt{4 ~ D(E_*)~t}$. This means that for energies well above $E_*$ the CR energy spectrum is a power law $f \propto N_{SNR}(E)/D(E)^{3/2} \propto E^{-\alpha-3\delta/2}$, while a sharp low energy cutoff suppress the spectrum below $E_*$. Then, for photon energies above $E_{\gamma}^* \approx 0.1 E_*$, the neutral pion decay gamma--ray spectrum from the cloud exhibit the same spectral shape as the CR spectrum, while below that energy one expects to see a spectrum close to the one produced by a monoenergetic distribution of protons, which is flat: $\propto E_{\gamma}^0$ \cite{dermer,kelner,kamae}.

To describe a continuous injection of CRs, the appropriate injection term is: $Q = Q_{SNR}(E) \delta(R)$, with $Q_{SNR} \propto E^{-\alpha}$ as the injection spectrum. Assuming that the injection lasts for a sufficiently long time, the steady state solution of Eq.~\ref{eq:prop} can be found and reads:
\begin{equation}
f(E,R) = \frac{Q_{SNR}(E)}{4 \pi D(E) R}
\end{equation}
Thus, in this case CRs are not uniformly distributed but their intensity decreases with the distance from the source as $\propto 1/R$ and their spectrum is $\propto E^{-\alpha-\delta}$. The steady state solution is applicable when the duration of the injection is much longer than the diffusion time of CRs $t \approx R^2/(6~D)$. Considerations similar to the ones made for the impulsive source can be made to infer the spectral shape of the gamma rays produced by the runaway CRs.

It must be remembered that, when computing the gamma--ray emission from a MC located in the vicinity of a CR accelerator, also the (unavoidable!) contribution from the CR background has to be considered. This would add a soft and steady component to the gamma--ray emission which would often dominate the emission in the GeV energy range.

In \cite{gabici2009}, the gamma--ray spectrum due to proton--proton interactions in a MC was computed for different distances from a nearby SNR and for different times after the supernova explosion. In these work, the SNR is assumed to inject $\approx 3 \times 10^{50}$~erg in form of CRs, with a spectrum $\propto E^{-2}$. To mimic an energy dependent escape of particles from the SNR, particles of energy $E$ are assumed to escape after a time $t_{esc} \propto E^{-\omega}$, with $\omega$ chosen in a way to have particles with energy $4 \times 10^{15}$~eV (the CR knee) escaping the SNR at the beginning of the Sedov phase, and particles with energy 1 GeV escaping at the end of the Sedov phase. Thus, the CR distribution around the SNR can still be described by Eq.~\ref{eq:impulsive} with a modified definition of the diffusion distance: $R_d = \sqrt{4 ~ D(E) ~ (t-t_{esc}(E))}$, that takes into account that CRs with different energies are released at different times.

The main finding is that concave gamma--ray spectra may be produced in a MC located in proximity of a SNR, as the result of the decay of neutral pions produced in CR interactions. Such concavity reflects the shape of the underlying CR spectrum, which consists of the superposition of two components: the galactic CR background, characterized by a steep spectrum, and the CRs coming from the nearby SNR, which exhibit a hard spectrum.
With this respect, the distance between the SNR and the MC $d_{SNR/MC}$ plays a crucial role.
This is because, the larger the distance between the SNR and the cloud, the lower the level of the CR flux coming from the SNR.
Moreover, also the time evolution of the emission from a cloud changes with $d_{SNR/MC}$ since the time it takes a particle with given energy to cover such a distance scales as $t \sim d_{SNR/MC}^2/D$, where $D$ is the diffusion coefficient. 

\begin{figure}[t!]
 \centering
 \includegraphics[width=0.32\textwidth]{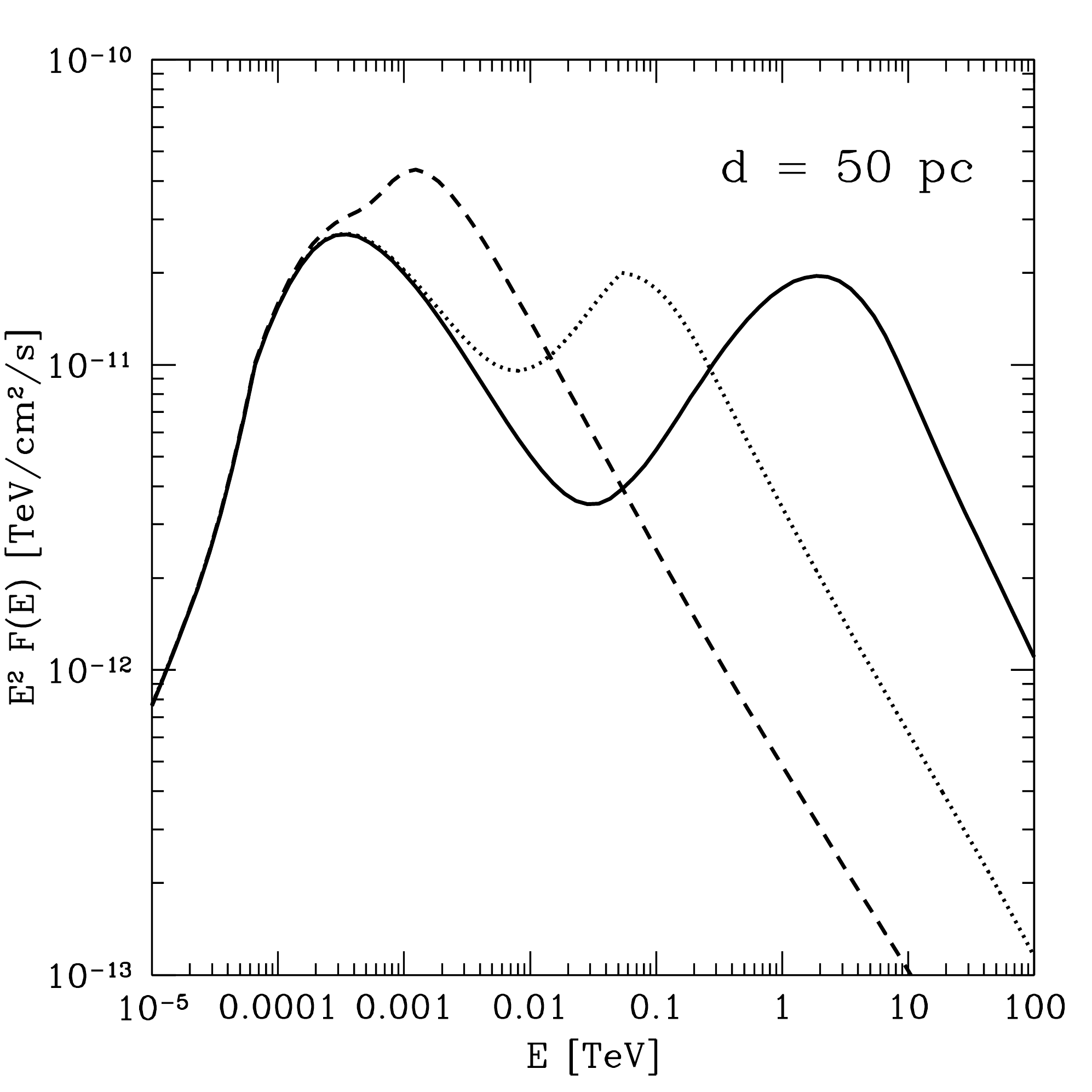}      
  \includegraphics[width=0.32\textwidth]{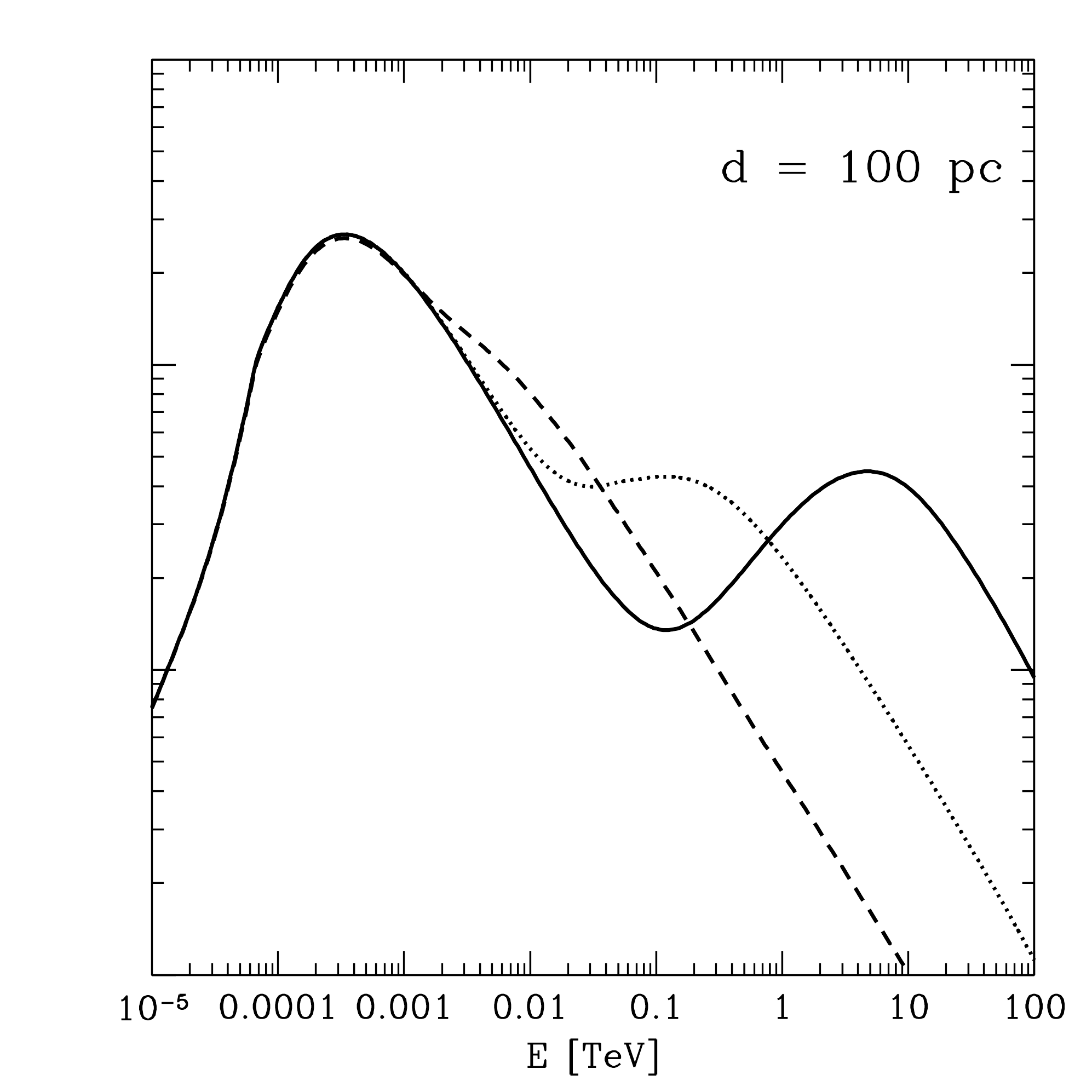}
  \includegraphics[width=0.32\textwidth]{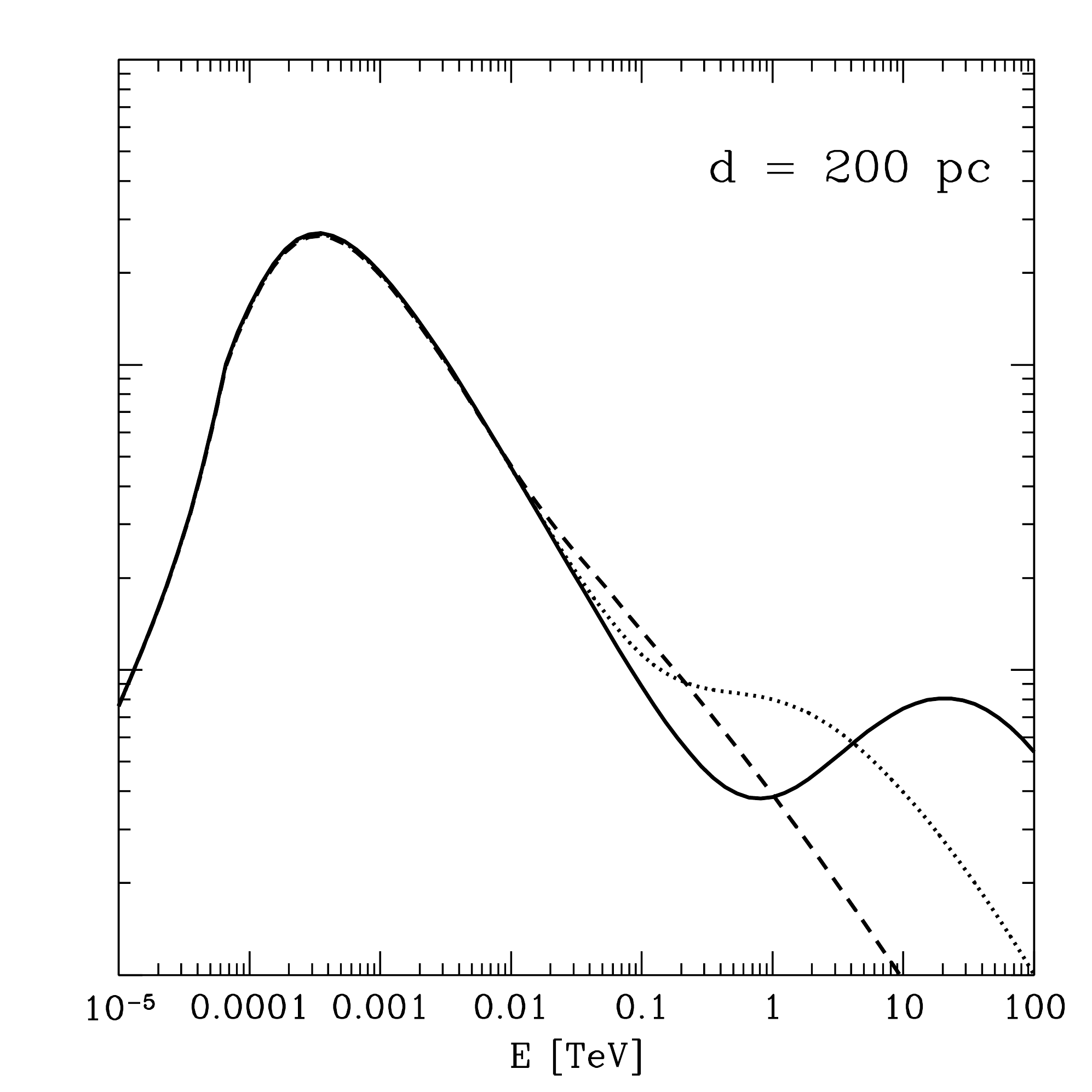}
  \caption{Total gamma ray emission from a molecular cloud of mass $10^5 M_{\odot}$ located at a distance of 1 kpc. The distance between the MC and the SNR is
50, 100 and 200 pc for left, centre and right panel, respectively. The solid, dotted,
and dashed lines refers to the emission at a time 2000, 8000, and 32000 years after
the explosion.}
  \label{fig:Vshaped}
\end{figure}

In Fig.~\ref{fig:Vshaped} the total gamma ray spectrum from a MC is shown as a function of the distance between the SNR and the cloud. The cloud mass is $10^5 M_{\odot}$ and the distance from the SNR is 50, 100 and 200 pc for the left, central and right panel, respectively.
The solid, dotted, and dashed lines refer to the emission for 2000, 8000, and 32000 years after the supernova explosion.
It is evident from Fig.~\ref{fig:Vshaped} that a great variety of gamma--ray spectra can be produced.
In almost the entirety of the cases considered, the gamma ray--emission is characterized by the presence of two pronounced peaks.
The low energy peak, located in the GeV domain is steady in time and it is the result of the decay of neutral pions produced in hadronic interactions of background CRs in the dense intercloud gas. The high energy peak is the result of hadronic interactions of CRs coming from the nearby SNR, and thus it is moving in time to lower and lower energies, as CRs with lower energies can reach the MC at later times.
Both the relative intensity and position of the two peaks depend on the distance between the SNR and the cloud.
Interestingly, the GeV emission from the cloud is affected by the presence of the nearby SNR only at late times after the explosion and only if the distance from the SNR is comparable or smaller than $\approx$ 50 pc (see Fig.~\ref{fig:Vshaped}, left panel).
In all the other cases the GeV emission is always the result of the interactions of background CRs and thus, at least in this case, observations of molecular clouds in the GeV gamma ray domain cannot be used to infer the presence of a CR accelerator located at a distance greater that $\approx$ 50 pc from the cloud. 

Besides the results described above, a large amount of theoretical and phenomenological work has been carried out by several research groups in order to describe the gamma--ray emission from SNR/MC associations, and the reader is referred to e.g. \cite{diego,sabrinaRXJ,donescape} for a list of relevant publications.

In the next Sections, the results derived here will be applied to two scenarios. In Sec.~\ref{sec:PeV} it will be shown how the detection of multi--TeV gamma rays from MCs can be used to identify the location of CR PeVatrons in the Galaxy, while in Sec.~\ref{sec:W28} the gamma--ray emission detected from the MCs in the vicinity of the SNR W28 will be modeled and a value for the CR diffusion coefficient will be derived as an output of the modeling.

\section{An application: gamma--ray emission from molecular clouds and the origin of galactic cosmic rays up to the knee}
\label{sec:PeV}

As said in the Introduction, if SNRs are the sources of galactic CRs, they have to accelerate them all the way up to the CR knee, at $\sim 4 \times 10^{15} {\rm eV} = 4 ~ {\rm PeV}$. In other words, they must act as CR PeVatrons. If this is indeed the case, SNRs are expected to emit gamma--rays due to hadronic interactions between the accelerated CRs and the interstellar medium swept up by the shock wave with a spectrum extending up to photon energies of hundreds of TeV.

The detection of a SNR in gamma rays with energies up to hundreds of TeV would constitute a decisive and unambiguous indication of acceleration of PeV protons. Because of the Klein-Nishina effect (e.g. \cite{gould}) the efficiency of 
inverse Compton scattering 
in this energy band is dramatically reduced. Therefore unlike 
other energy intervals, the interpretation of 
gamma-ray observations at these energies is  
free of confusion and reduces to the only possible 
mechanism - decay of secondary $\pi^0$-meson. 
Although the potential of the current ground-based 
instruments for detection of such energetic 
gamma-rays is limited, it is expected that 
the next generation arrays of imaging Cherenkov telescopes, exploring a broad energy region extending up to the multi--TeV energy range will become 
powerful tools for this kind of studies. 

It should be noted that the number of SNRs currently bright in $> 10$ TeV gamma rays 
is expected to be rather limited. Multi--PeV protons can be 
accelerated only during a relatively short period of the SNR evolution, namely, at the end of the free--expansion phase/beginning of the Sedov phase, when the shock velocity is high enough to allow a sufficiently high acceleration rate. When the SNR enters the Sedov phase, the shock gradually slows down and correspondingly the maximum energy of the particles that can be confined within the SNR decreases. This determines the escape of the most energetic particles from the SNR \cite{ptuskinzirakashvili}. Thus, unless our theoretical understanding of particle acceleration at SNR is completely wrong, we should expect an energy  spectrum of CR inside the SNR approaching  
PeV energies only at the beginning of the Sedov phase, typically 
for a time $\lesssim 1000$ years. 
When the remnant enters the Sedov phase, the high energy cutoffs in the spectra of 
both protons and gamma rays gradually moves to lower energies, while the highest energy particles 
leave the remnant \cite{ptuskinzirakashvili}. 

Here we suggest to search for multi--TeV gamma-rays 
generated by the CRs that escape the SNR. 
A molecular cloud located close to the SNR can provide an effective 
target for production of gamma--rays. 
The highest energy particles ($\sim$ few PeV) escape the shell 
first. Moreover, they diffuse in the interstellar medium 
faster than low energy particles. Therefore they arrive first to the cloud,
producing there gamma rays with very hard energy spectra.  
Note that an association of SNRs with clouds is naturally 
expected, especially in star forming regions \cite{snobs}. 
The duration of the gamma-ray emission in this case is determined by the time 
of propagation of CRs from the SNR to the cloud, which in turn depends on the value of the CR diffusion coefficient in the vicinity of the SNR.
It is a very well known fact that the CR diffusion coefficient at specific locations in the Galaxy is very poorly constrained from observations, and theoretical predictions are still far from giving solid and reliable estimates for this quantity.
However, it is often believed that the CR diffusion coefficient in the vicinity of CR sources  might be suppressed with respect to the average galactic one due to CR streaming instability \cite{wentzel,cesarsky}.
This CR-induced instability would increase the magnetic turbulence and in turn suppress the diffusion of CRs themselves.
Therefore the gamma-ray emission of the cloud may last much longer
than the emission of the SNR itself. This makes the detection of 
delayed gamma-ray signal from clouds more probable.
The detection of these multi--TeV gamma-rays from 
nearby clouds would thus indicate that the nearby SNR 
in the past was acting as an effective CR PeVatron.

\begin{figure}[t!]
 \centering
 \includegraphics[width=0.6\textwidth]{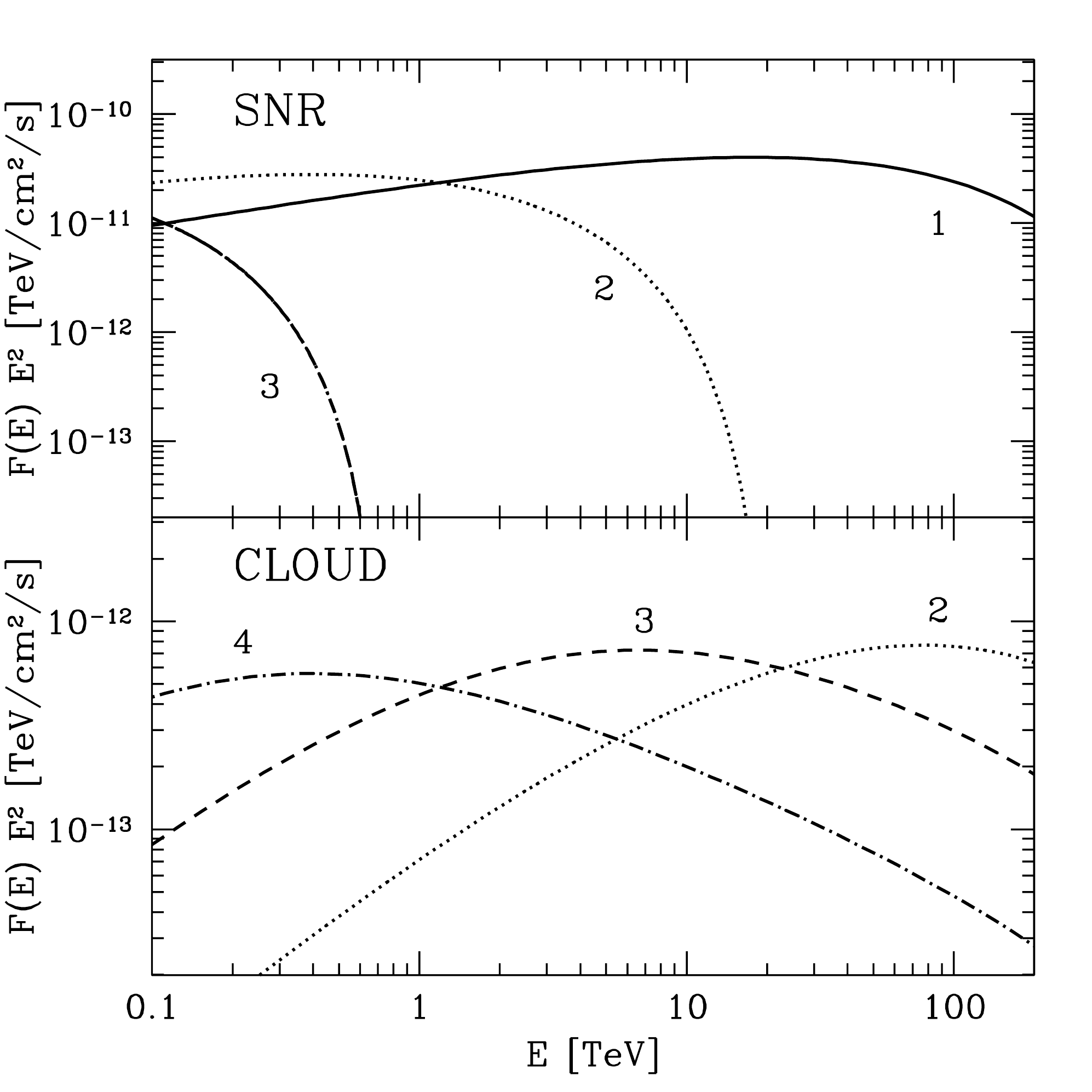}      
\caption{Gamma--ray spectra from the SNR (TOP) and from a cloud of $10^4 M_{\odot}$
located 100 pc away from the SNR (BOTTOM). The distance is 1 kpc. Curves refer
to different times after the explosion: 400 (curve 1), 2000 (2), 8000 (3), 3200 (4)
yr. Figure from \cite{gabiciPeV}}
  \label{fig:1-PeV}
\end{figure}

The top panel of Fig. \ref{fig:1-PeV} \cite{gabiciPeV} shows the predicted energy spectrum of gamma-ray emission from a SNR produced by interactions of accelerated protons with ambient medium, calculated for typical parameters characterizing SNRs: an explosion energy equal to $10^{51}$~erg, an ambient density of 1 cm$^{-3}$ and and initial shock velocity of $10^9$~cm/s. The bottom panel shows the emission from a cloud of mass $M_{cl} = 10^4 M_{\odot}$ located at a distance $d_{cl} = 100$ pc away from the SNR. 
The distance of the SNR is assumed $D = 1$ kpc and different curves refer to different times after the supernova explosion. The efficiency of CR acceleration at the SNR shock is regulated by the parameter $\xi_{CR}$ (the ratio between the CR pressure at the shock to the shock ram pressure), which is assumed to be equal to 0.3 and constant during the SNR evolution. Finally, we assume a value of the diffusion coefficient equal to $D_{ISM} = 3 \times 10^{29} (E/1~ PeV)^{0.5} {\rm cm}^2/{\rm s} $, which is significantly suppressed with respect to the extrapolation at PeV energies of the average galactic one, which is measured at much lower energies.  

Early in the Sedov phase (curve 1, 400 yr after the explosion), the gamma-ray spectrum from the SNR is hard and extends up to $\gtrsim 100$ TeV, revealing the acceleration of PeV particles. 
Conversely, the gamma-ray flux from the cloud is extremely weak, because for the epoch of 400 yr after the explosion CRs do not have sufficient time to reach the cloud. The emission of $\gtrsim 100~ {\rm TeV}$ photons from the SNR lasts a few hundreds years, and after that the cutoff in the gamma-ray spectrum moves to lower energies (curves 2, 3 and 4 correspond to the epochs of $2 \times 10^3$, $8 \times 10^3$ and $3.2 \times 10^4$ yr after the explosion).
As time passes, CRs finally reach the cloud and produce there gamma rays when interacting with the dense cloud environment. This makes the cloud an effective multi--TeV gamma-ray emitter, with a flux at the sensitivity level of next generation Cherenkov telescopes operating in that energy range. 
As lower and lower energy particles reach the cloud, the peak of the gamma-ray emission shifts accordingly towards lower energies, first $\approx$~TeV, then $\approx$~GeV, at flux levels which can be probed by ground based instruments and by Fermi. 

As discussed in the previous Section, the shape of the gamma--ray spectrum is naturally explained as follows: at a time $t$, only particles with energy above $E_{*}$, given by $d_{SNR/MC} \approx \sqrt{6 D_{ISM}(E_{*}) t}$, reach the cloud. Thus the CR spectrum inside the cloud has a sharp low energy cutoff at $E_{*}$.
The corresponding gamma-ray spectrum exhibits a prominent peak at the energy $\approx 0.1 E_{*}$.

The multi-TeV hadronic gamma-ray emission from the cloud is significantly weaker than the one from the SNR, but its detection might be easier because of its longer duration ($\lesssim 10^4$ yr versus few hundreds years). Moreover, 
the leptonic contribution to the cloud emission is likely to be negligible.
Electrons accelerated at the SNR cannot reach the cloud because they remain confined in the SNR due to severe synchrotron losses. Secondary electrons can be produced in the cloud, but they cool mainly via synchrotron emission in the cloud magnetic field $\sim 10 \div 100 ~ \mu G$ \cite{crutcher}. This makes the production of $\gtrsim$ TeV gamma rays due to inverse Compton scattering and non--thermal Bremsstrahlung negligible.

To conclude, the acceleration of CRs up to the knee in SNRs can be unambiguously revealed by means of observations of multi-TeV gamma rays and neutrinos coming from the SNR and nearby MCs. The emission from the clouds is weaker than the one from the SNR, but may last much longer, depending on the actual value of the diffusion coefficient, and this might significantly enhance the probability of detection.
Gamma rays are emitted with fluxes detectable by currently operating and forthcoming instruments.  
Since the gamma-ray spectra from clouds are extremely  hard, gamma-ray telescopes operating up to very high energies ($\gtrsim 10$~TeV), like the Cherenkov Telescope Array, would be the best instruments for this kind of study. 

\section{Another application: constraining the diffusion coefficient in the region surrounding the supernova remnant W28}
\label{sec:W28}

W28 is an old SNR in its radiative phase of evolution, located in a region rich of dense molecular gas with average density $\gtrsim 5~{\rm cm^{-3}}$. At an estimated distance of $\approx 2~{\rm kpc}$ the SNR shock radius is $\approx 12~{\rm pc}$ and its velocity $\approx 80~{\rm km/s}$ (see e.g. \cite{rho}). By using the dynamical model from \cite{cioffi} and assuming that the mass of the supernova ejecta is $\sim 1.4 ~ M_{\odot}$, it is possible to infer the supernova explosion energy ($E_{SN} \approx 0.4 \times 10^{51} {\rm erg}$), initial velocity ($\approx 5500~{\rm km/s}$), and age ($t_{age} \approx 4.4 \times 10^4 {\rm yr}$).

Gamma ray emission has been detected from the region surrounding W28 both at TeV \cite{W28HESS} and GeV energies \cite{W28FERMI,W28AGILE}, by {\it HESS}, {\it FERMI}, and {\it AGILE}, respectively. The TeV emission correlates quite well with the position of three massive molecular clouds, one of which is interacting with the north-eastern part of the shell (and corresponds to the TeV source HESS J1801-233), and the other two being located to the south of the SNR (TeV sources HESS J1800-240 A and B) . The masses of these clouds can be estimated from CO measurements and result in $\approx 5$, $6$, and $4 \times 10^4 M_{\odot}$, respectively, and their projected distances from the centre of the SNR are $\approx$ 12, 20, and 20 pc, respectively \cite{W28HESS}. The GeV emission roughly mimics the TeV one, except for the fact that no significant emission is detected at the position of HESS J1800-240 A. 

Here, we investigate the possibility that the gamma ray emission from the W28 region could be the result of hadronic interactions of CRs that have been accelerated in the past at the SNR shock and then escaped in the surrounding medium \cite{gabici2010}\footnote{This scenario has been described in a number of recent papers \cite{fujita,ohira,li,yan}.}. To do so, we follow the approach described in Sec.~\ref{sec:SNRMC} and in \cite{gabici2009}, which we briefly summarize below. 

For each particle energy $E$ we solve the diffusion equation for CRs escaping the SNR. For simplicity we treat the SNR as a point like source of CRs and we consider an isotropic and homogeneous diffusion coefficient: $D \propto E^{\delta}$. 
A value of $\delta = 0.5$ is found to provide a good fit to data (see below), though reasonably good fits can be obtained also for values in the range $\delta = 0.3 - 0.7$. The solution of the diffusion equation gives the spatial distribution of CRs around the source $f_{CR}$, which is roughly constant up to a distance equal to the diffusion radius $R_d = \sqrt{4 ~ D ~ t_{diff}}$, and given by $f_{CR} \propto \eta E_{SN}/R_d^3$, where $\eta$ is the fraction of the supernova explosion energy converted into CRs, and $t_{diff}$ is the time elapsed since CRs with energy $E$ escaped the SNR. For distances much larger than $R_d$ the CR spatial distribution falls like $f_{CR} \propto \exp(-(R/R_d)^2)$, as expected for diffusion (see Eq.~\ref{eq:impulsive}). Following the approach described in \cite{gabici2009}, we assume that CRs with energy 5 PeV (1 GeV) escape the SNR at the beginning (end) of the Sedov phase, at a time $\sim 250 ~ {\rm yr}$ ($\sim 1.2 \times 10^4 ~{\rm yr}$) after the explosion, and that the time integrated CR spectrum injected in the interstellar medium is $\propto E^{-2}$. In this scenario, particles with lower and lower energies are released gradually in the interstellar medium \cite{ptuskinzirakashvili}, and we parametrize the escape time as: $t_{esc} \propto E^{-\alpha}$ which, during the Sedov phase, can also be written as $R_s \propto E^{-2 \alpha /5}$, where $R_s$ is the shock radius at time $t_{esc}$ and $\alpha \sim 4$. From this it follows that the assumption of point like CR source is good for high energy CRs only ($\sim$ TeV or above), when $R_s$ is small, but it becomes a rough approximation at significantly lower energies. This is because low energy particles are believed to be released later in time, when the SNR shock radius is large (i.e. non negligible when compared to $R_d$). 

We now provide a simplified argument to show how we can attempt to constrain the diffusion coefficient by using the TeV gamma ray observations of the MCs in the W28 region. The time elapsed since CRs with a given energy escaped the SNR can be written as: $t_{diff} = t_{age} - t_{esc}$. However, for CRs with energies above 1~TeV (the ones responsible for the emission detected by HESS) we may assume $t_{esc} << t_{age}$ (i.e. high energy CRs are released when the SNR is much younger than it is now) and thus $t_{diff} \sim t_{age}$. Thus, the diffusion radius reduces to $R_d \sim \sqrt{4 ~ D ~ t_{age}}$. We recall that within the diffusion radius the spatial distribution of CRs, $f_{CR}$, is roughly constant, and proportional to $\eta E_{SN}/R_d^3$. On the other hand, the observed gamma ray flux from each one of the MCs is: $F_{\gamma} \propto f_{CR} M_{cl}/d^2$, where $M_{cl}$ is the mass of the MC and $d \approx 2$~kpc is the distance of the system. Note that in this expression $F_{\gamma}$ is calculated at a photon energy $E_{\gamma}$, while $f_{CR}$ is calculated at a CR energy $E_{CR} \sim 10 \times E_{\gamma}$, to account for the inelasticity of proton-proton interactions. By using the definitions of $f_{CR}$ and $R_d$ we can finally write the approximate equation, valid within a distance $R_d$ from the SNR:
$$
F_{\gamma} \propto \frac{\eta ~ E_{SN}}{(D ~ t_{age})^{3/2}} \left( \frac{M_{cl}}{d^2} \right) .
$$

Estimates can be obtained for all the physical quantities in the equation except for the CR acceleration efficiency $\eta$ and the local diffusion coefficient $D$. By fitting the TeV data we can thus attempt to constrain, within the uncertainties given by the errors on the other measured quantities (namely, $E_{SN}$, $t_{age}$, $M_{cl}$, and $d$) and by the assumptions made (e.g. the CR injection spectrum is assumed to be $E^{-2}$, while the energy dependence of $D$ is assumed to scale as a power law of index $\delta = 0.5$), a combination of these two parameters (namely $\eta/D^{3/2}$). The fact that the MCs have to be located within a distance $R_d$ from the SNR can be verified a posteriori. Given all the uncertainties above, our results have to be interpreted as a proof of concept of the fact that gamma ray observations of SNR/MC associations can serve as tools to estimate the CR diffusion coefficient. More detection of SNR/MC associations are needed in order to check whether the scenario described here applies to a whole class of objects and not only to a test-case as W28. Future observations from the Cherenkov Telescope Array will most likely solve this issue.

\begin{figure}[t!]
 \centering
 \includegraphics[width=0.6\textwidth]{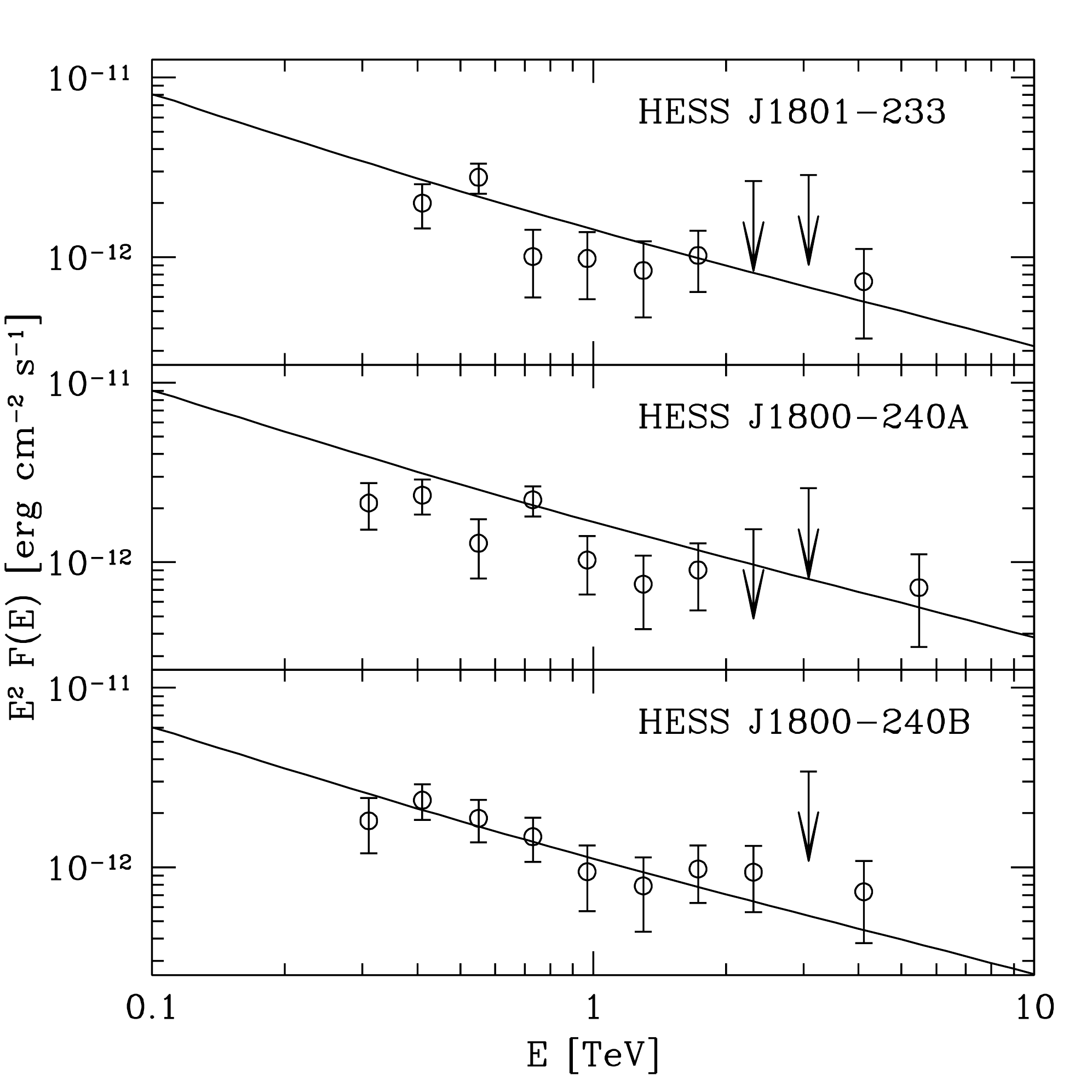}      
  \caption{Simultaneous fit to the three TeV sources detected by HESS in the W28 region. Gamma ray spectra have been calculated by using the parameterizations by \cite{kamae}, where a multiplicative factor of 1.5 has been applied to account for the contribution to the emission from nuclei heavier than H both in CRs and in the interstellar medium. Figure from \cite{gabici2010}}
  \label{fig1}
\end{figure}

Fig. \ref{fig1} shows a fit to the {\it HESS} data for the three massive MCs in the W28 region. A simultaneous fit to all the three MCs is obtained by fixing a value for  $\eta/D^{3/2}$, which implies that the diffusion coefficient of particle with energy $3~{\rm TeV}$ (these are the particles that produce most of the emission observed by {\it HESS}) is:
\begin{equation}
D(3~{\rm TeV}) \approx 5 \times 10^{27} ~ \left(\frac{\eta}{0.1}\right)^{2/3} ~ {\rm cm^2/s} ~ .
\end{equation}
This value is significantly smaller (more than an order of magnitude) than the one normally adopted to describe the diffusion of $\sim$~TeV CRs in the galactic disk, which is $\approx 10^{29}~{\rm cm^2/s}$. This result remains valid (i.e. a suppression of the diffusion coefficient is indeed needed to fit data) even if a different value of the parameter $\delta$ is assumed, within the range 0.3...0.7 compatible with CR data. 
As an example, an acceleration efficiency $\eta = 30\%$ corresponds to a CR diffusion coefficient of $D \sim 10^{28}~{\rm cm^2/s}$, which in turn gives a diffusion distance for 3~TeV particles of $R_d \approx 80~{\rm pc}$. This means that the results in Fig.~\ref{fig1} are valid if the physical (not projected) distances between the MCs and the SNRs do not significantly exceed $R_d$.  Small values of the diffusion coefficient have been also proposed by \cite{W28AGILE,fujita,li,yan}. Note that, since we are considering gamma rays in a quite narrow (about one order of magnitude) energy band around $\approx 1$~TeV, we can actually constrain the diffusion coefficient of CRs with energy $\approx 3 - 30$~TeV, and we cannot say much about the energy dependence of the diffusion coefficient 
on a broad energy interval. 

\begin{figure}[t!]
 \centering
 \includegraphics[width=0.32\textwidth]{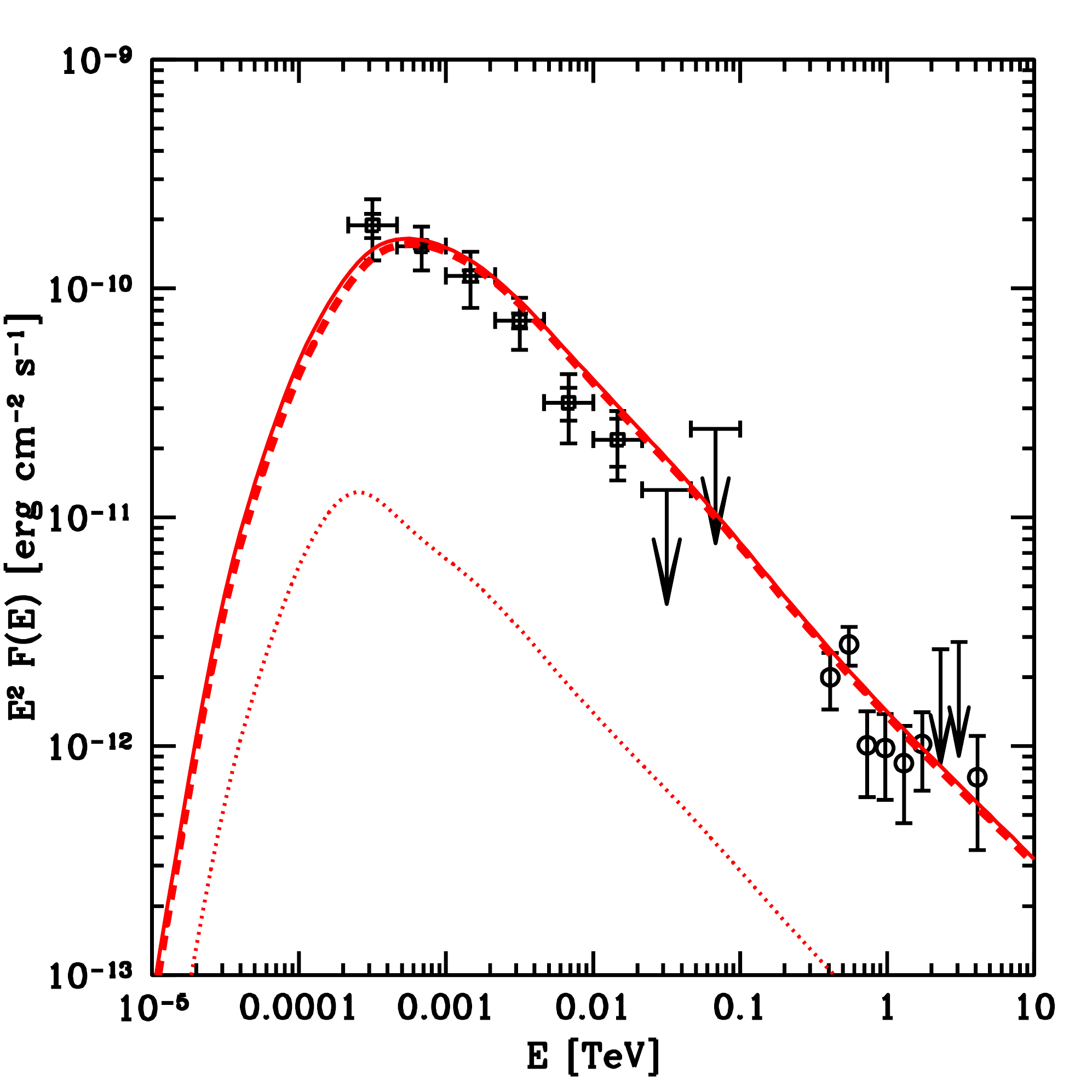}   
 \includegraphics[width=0.32\textwidth]{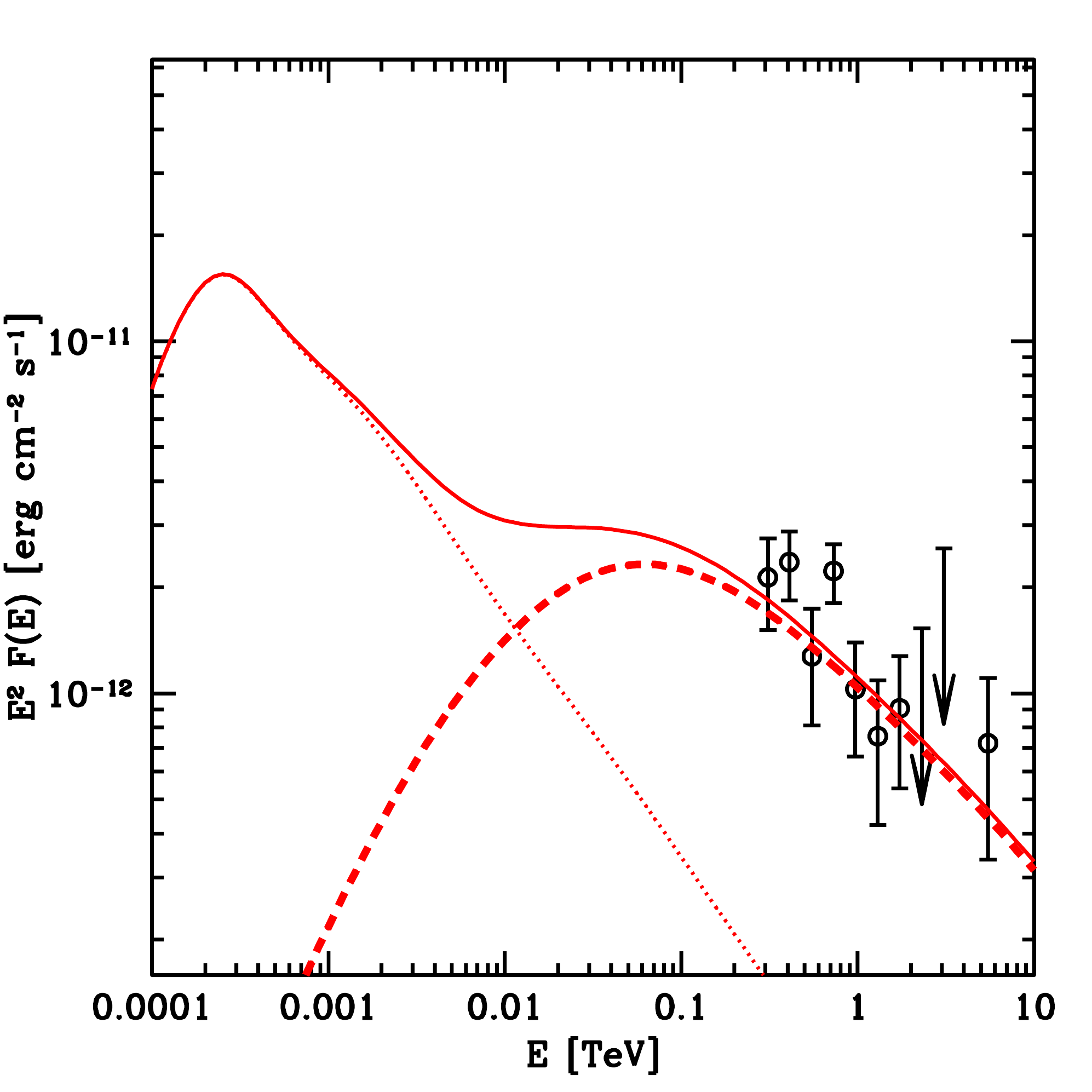}   
 \includegraphics[width=0.32\textwidth]{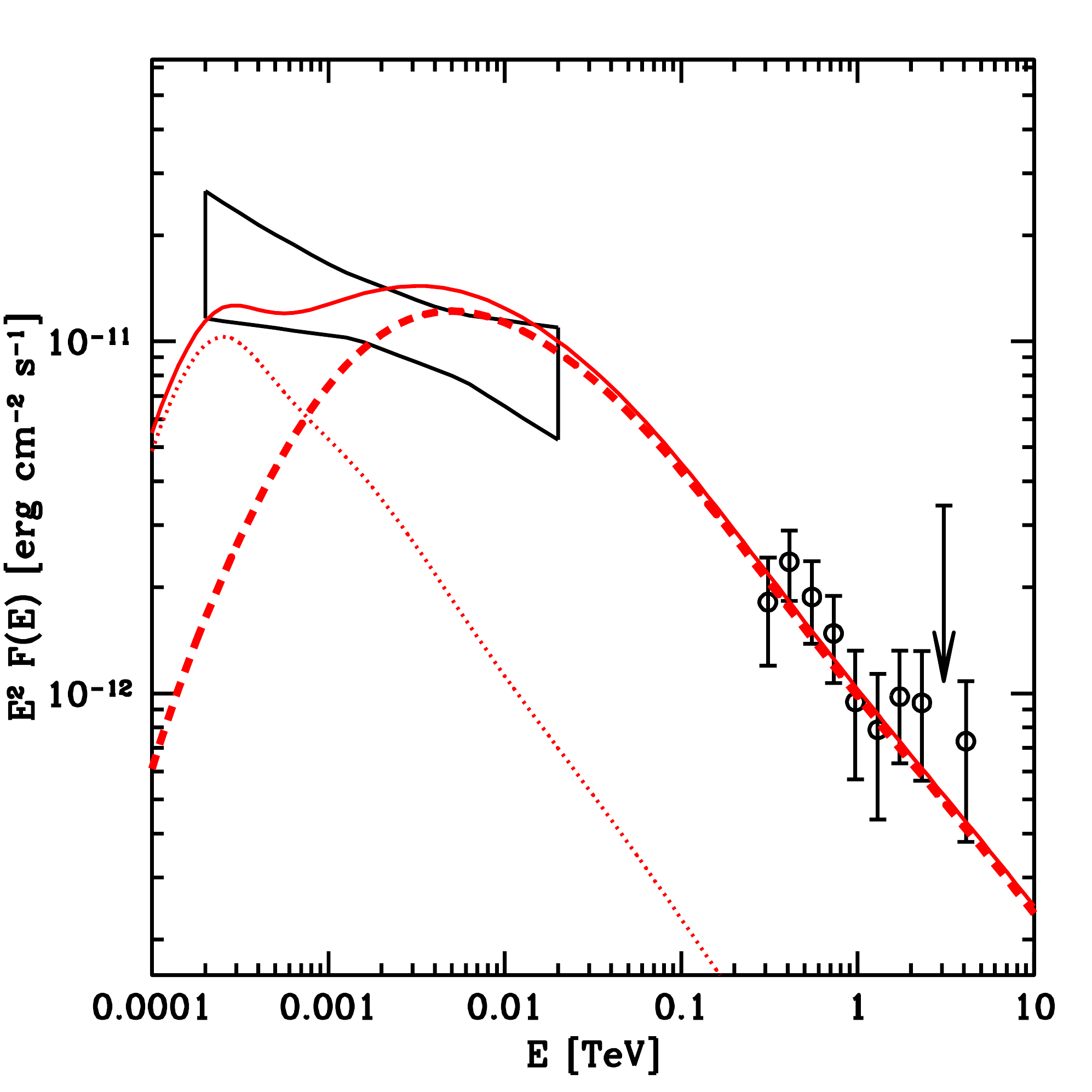}   
  \caption{Broad band fit to the gamma ray emission detected by {\it FERMI} and {\it HESS} from the sources HESS J1801-233, HESS J1800-240 A and B (left to right). Dashed lines represent the contribution to the gamma ray emission from CRs that escaped W28, dotted lines show the contribution from the CR galactic background, and solid lines the total emission. Distances to the SNR centre are 12, 65, and 32 pc (left to right). {\it FERMI} and {\it HESS} data points are plotted in black. No GeV emission has been detected from HESS J1800-240 A. Figure from \cite{gabici2010}.}
  \label{fig2}
\end{figure}

In principle, observations by {\it FERMI} and {\it AGILE} might be used to constrain the diffusion coefficient down to GeV particle energies. However, in this energy band the uncertainties are more severe because of the following reasons: {\it i)} low energy CRs are believed to be released late in time, when the SNR shock is large, and thus the assumption of point-like source is probably not well justified (see \cite{ohira} for a model that takes into account the finite size of the SNR) ; {\it ii)} for the same reason, we can no longer assume that $t_{diff} \sim t_{age}$, as we did for high energy CRs. In other words, we need to know the exact time at which CRs with a given energy escape the SNR. Though some promising theoretical studies exist (see \cite{gabiciescape} and references therein), our knowledge of the escape time of CRs from SNRs is still quite limited.

Fig.~\ref{fig2} shows a fit to the broad band gamma ray spectrum measured from {\it FERMI} and {\it HESS}. The three panels refers to (left to right) the sources HESS J1801-233, HESS J1800-240 A and B, respectively. Dashed lines represent the contribution to the emission from CRs that escaped from W28, dotted lines the contribution from background CRs, and solid lines the total emission. Since {\it FERMI} data refers to the emission after background subtraction, dashed lines have to be compared with data points. The (often non-trivial) background subtraction issue might add another source of uncertainty in the comparison between data and predictions. An acceleration efficiency $\eta = 30\%$ and a diffusion coefficient $D = 10^{28}~{\rm cm^2/s}$ at 3~TeV have been assumed, while the distance from the SNR centre is assumed to be (left to right) 12, 65, and 32 pc. Keeping in mind all the above mentioned caveats, it is encouraging to see that a qualitative agreement exists between data and predictions also in the GeV band.

Summarizing, we investigated the possibility that the gamma ray emission detected from the MCs in the region of the SNR W28 is produced by CRs that escaped the SNR. This interpretation requires the CR diffusion coefficient in that region to be significantly suppressed with respect to the average galactic one. Such suppression might be the result of an enhancement in the magnetic turbulence due to CR streaming away from the SNR.

\section{Conclusions and future perspectives}

In this paper, it has been shown how gamma--ray observations of MCs can be used to identify the location of the sources of galactic CRs and to constrain the CR diffusion properties close to the sources. Despite encouraging results from both observations and theory, further work is needed to reach a conclusive evidence in favor (or against) the SNR paradigm for the origin of CRs. Hopefully, this will come with the advent of the next generation of gamma--ray instruments, such as CTA.

The main limitations of the approach presented here are probably connected with the oversimplified assumptions made to describe the way in which particles escape from SNR shocks, which is still not well understood (see \cite{gabiciescape} and references therein for a discussion) and the way in which they diffuse in the interstellar medium.

Of great relevance is the fact that the presence of a MC interacting with the SNR shock can, one one side, amplify the gamma--ray emission from neutral pion decay \cite{adv}, but also influence and modify the acceleration mechanism of particles at shocks \cite{lukeneutrals,brianneutrals,crushedyas}, and even have important effects on their escape \cite{malkovbreak}. All these aspects need to be further investigated.

Moreover, the assumption of isotropic and homogeneous diffusion of CRs close to their sources is certainly an excessive oversimplification. In fact, CR are expected to diffuse preferentially along, parallel to, the magnetic field lines, the perpendicular transport being determined mainly by the wandering of the magnetic field line CRs are attached to \cite{parker,fabien}. In adition to that, the CR diffusion along the magnetic field lines is most likely a non--linear process, where CRs themselves generate the magnetic turbulence needed to confine them. Some preliminary work including the effects of anisotropic CR diffusion have been recently published \cite{plesser,michael,malkovfelix,lara}, and this promises to become one of the most important developments in this field.

Finally, recent measurements of the CR ionization rate in a MC interacting with the shock of the SNR W51C  have been presented \cite{cecilia}. These data reveal an enhancement in the CR ionization rate of about 2 orders of magnitude with respect to standard values. Such an enhancement might be interpreted as the result of the presence of CRs accelerated at the SNR shock. The SNR/MC system has been also detected in TeV gamma rays \cite{julian}, and the emission is most likely hadronic. Thus, in this particular system CRs can be studied from $\approx$~MeV energies (the most relevant for the ionization of the gas) up to multi TeV energies. In the future, studies of this kind will shed light on the acceleration of CRs at shocks and on their escape over an unprecedented energy range.

\begin{acknowledgement}
I would like to thank the organizers of the Sant Cougat Forum for Astrophysics, Diego Torres and Olaf Reimer, for their invitation. I also acknowledge support from the EU [FP7--grant agr. n$^{\circ}$256464] and from ANR [JCJC Programme].
\end{acknowledgement}

\end{document}